\documentclass[12pt]{article}

\usepackage{a4,amsmath,amssymb,amsthm,amscd,cite,graphicx}
\usepackage{verbatim,numprint,siunitx,mathrsfs,esint,xcolor}
\usepackage{rotating}
\usepackage{hyperref} \hypersetup{colorlinks, linkcolor=red }
\usepackage{tikz}
\usetikzlibrary{intersections,calc,matrix,arrows, decorations.markings,shapes}
\usepackage{multirow}
\newtheorem{theorem}{Theorem}[section]
\newtheorem{mydef}[theorem]{Definition}
\newtheorem{prop}[theorem]{Proposition}

\def\Bbb{\mathbb} \def\BZ{\Bbb Z} \def\BR{\Bbb R} \def\BC{\Bbb C}
  \def\BH{\mathbb{H}}

\newcommand{\slhat}{{\widehat{sl(2)}}}
\newcommand{\hdelta}{{\widehat{\delta}}}
\catcode`@=11 \@addtoreset{equation}{section} \catcode`@=12

\begin{document}

\begin{titlepage}
\begin{flushright}
 June 2021 (v1)\\
\texttt{arXiv:2106.xxxxx [hep-th]}
\end{flushright}
\begin{center}
\textsf{\large $\slhat$ decomposition of denominator formulae of some BKM Lie superalgebras}\\[12pt]
Suresh Govindarajan$^{a}$ Mohammad Shabbir$^{b}$ and Sankaran Viswanath$^{b,}$\footnote{SV is partially supported by SERB-Matrics grant number MTR/2019/000071.} \\[4pt]
$^a$Department of Physics,
Indian Institute of Technology Madras ,
Chennai 600036 India\\
 $^b$ The Institute of Mathematical Sciences, HBNI, Chennai 600113 India 
\\[4pt]
Email: suresh@physics.iitm.ac.in, mshabbir@imsc.res.in, svis@imsc.res.in
\end{center}
\begin{abstract}
We study a family of Siegel modular forms that are constructed using Jacobi forms that arise in Umbral moonshine. All but one of them arise as the 
Weyl-Kac-Borcherds denominator formula of some Borcherds-Kac-Moody (BKM) Lie superalgebras. These Lie superalgebras have a $\slhat$ subalgebra which we use to study the Siegel modular forms. We show that the expansion of the Umbral Jacobi forms in terms of $\slhat$ characters leads to vector-valued modular forms. We obtain closed formulae  for these vector-valued modular forms. In the Lie algebraic context, the Fourier coefficients of these vector-valued modular forms are related to multiplicities of roots appearing on the sum side of the Weyl-Kac-Borcherds denominator formulae. 
\end{abstract}
\end{titlepage}

\section{Introduction}

The use of automorphic forms in the study of Lie algebras first arose in the work of Macdonald. He associated Jacobi forms with denominator formulae for affine Lie algebras and determined the multiplicity of positive roots using this connection\cite{MacDonald:1971,Macdonald:1981}. For instance, for $\slhat$, the product side of the denominator formula is given by\footnote{In this paper, we follow the notation $q=\exp(2\pi i \tau)$, $r=\exp(2\pi i z)$ and $s=\exp(2\pi i \tau')$.}
\begin{equation}
s^{1/2}\,\vartheta_1(\tau,z) = s^{1/2}q^{1/8}r^{1/2}\prod_{m=1}^\infty (1-q^m)(1-q^m r)(1-q^{m-1} r^{-1}) \ .
\end{equation}
Here we make the identifications: $e^{-\alpha_1}\sim q r$, $e^{-\alpha_2}\sim r^{-1}$ and $e^{-\delta}=e^{-\alpha_1-\alpha_2}\sim q $. The right hand side of the above equation then reads (with $e^{-\varrho}\sim  s^{1/2}q^{1/8}r^{1/2}$ )
\[
e^{-\varrho} \prod_{\alpha\in L_+} (1-e^{-\alpha})\ ,
\]
where $\varrho$ is the Weyl vector and the set of positive roots $L_+$ is
\[
L_+ =\{m \delta, (m-1)\delta + \alpha_1, (m-1)\delta +\alpha_2 ~|~ m\in \BZ_{>0}\}\ .
\]
This product formula shows that one needs to include imaginary roots, of type $m\delta$, in the the set of positive roots. 

Borcherds extended this by including situations where imaginary simple roots also appear\cite{Borcherds1988}.  This leads to modifications on the sum side of the denominator formulae to account for such simple roots. Today, such Lie algebras are called Borcherds-Kac-Moody (BKM) Lie algebras.  Again, automorphic forms play an important role in determining the sum and product side of the denominator formulae which we call the Weyl-Kac-Borcherds denominator formulae. Schematically, one has
\[
 \Delta=\sum_{w\in W}\text{det}(w) w \Big[ e^{-\varrho}\ T\Big]  = e^{-\varrho}\ \prod_{\alpha\in L_+} (1-e^{-\alpha})^{\text{mult}(\alpha)} \ .
\]
In the above formula, an automorphic form $\Delta$ is written as a sum and a product. Further, $T$ is the Borcherds correction factor defined in Eq. \eqref{eq:sumside},    $L_+$ corresponds to the set of positive roots and $\text{mult}(\alpha)$ is the multiplicity of the root $\alpha$. The addition of fermionic roots leads to superalgebras and we shall focus on superdenominator formulae where a similar correspondence described above goes through.

This paper studies the connection between five genus two Siegel modular forms (defined in Eq. \eqref{SMFdef})
\begin{equation}
\Delta_{k(N)}(\mathbf{Z})\ , \text{ for }N=1,2,3,4,6 \quad \text{ and }\quad k(N)= (6/N)-1\ ,
\end{equation}
and the BKM Lie superalgebras associated with five rank-three Cartan matrices $A^{(N)}$. The Cartan matrices are the inner products of the bosonic real simple roots and are of rank three. The Cartan matrices are given by
\begin{equation}\label{CartanMatrix}
A^{(N)}= (a_{nm}),\quad \text{where } a_{nm}= 2 - \frac{4}{N-4}(\lambda_N^{n-m} + \lambda_N^{m-n}-2)\ ,
\end{equation}
where $\lambda_N$ is any solution of the quadratic equation 
\begin{equation*}
\lambda^2 -(N-2)\lambda + 1 =0\ .
\end{equation*}
 For $N=1,2,3$, the matrices are finite-dimensional with the indices $n,m$ defined modulo $3,4,6$ respectively. For $N=4,6$, the matrices are infinite-dimensional with $m,n\in \BZ$. For $N=4$, the Cartan matrix has to be obtained as a limit $N\rightarrow4$ leading to $a_{nm}= 2 -4(n-m)^2$.

The square of these modular forms appear as the generating function of the refined counting of quarter-BPS states in certain CHL $\mathbb{Z}_N$ orbifolds of the heterotic string compactified on $T^6$\cite{Govindarajan:2010fu, Govindarajan:2019ezd}. Further, the root lattices associated with these Cartan matrices are closely related to the walls of marginal stability where quarter BPS states decay\cite{Sen:2007vb,Cheng:2008fc,Cheng:2008kt,Govindarajan:2009qt,Krishna:2010gc}.

Let $\mathfrak{g}(A^{(N)})$ denote the Kac-Moody algebra associated with the Cartan matrix $A^{(N)}$\cite{Kac1990}. 
For $N\neq 6$, the Siegel  modular form arises as the Weyl-Kac-Borcherds superdenominator formula for a Borcherds-Kac-Moody (BKM) Lie superalgebra that is a Borcherds extension of  $\mathfrak{g}(A^{(N)})$ by the addition of imaginary simple roots\cite{GritsenkoNikulinI,GritsenkoNikulinII}. Let us denote this BKM Lie superalgebra by $\mathcal{B}(A^{(N)})$. For $N=6$, let  $\mathcal{B}(A^{(6)})$ denote an as yet unknown Lie superalgebra whose WKB superdenominator formula is given by $\Delta_{k(6)}(\mathbf{Z})$.    All five Siegel modular forms admit a product formula of the form\cite{Govindarajan:2019ezd} (see Eq. \eqref{SMFdef})
\[
\Delta_{k(N)}(\mathbf{Z}) = e^{-2\pi i\text{Tr}(\varrho^{(N)}\,\mathbf{Z})}\ \prod_{\alpha\in L_+} (1-e^{-2\pi i\text{Tr}(\alpha\,\mathbf{Z})})^{m(\alpha)}\ ,
\]
where $L_+$ is the  set of positive roots implicitly determined by the product formula and $\varrho^{(N)}$ is the Weyl vector that satisfies $\langle \varrho^{(N)}, \alpha\rangle=-1$ for all simple real roots $\alpha$. The multiplicities $m(\alpha)$ are determined by the Fourier-Jacobi coefficients of a weight zero, index $N$, Jacobi form that appears in the context of umbral moonshine\cite{Cheng:2012tq}.

These BKM Lie superalgebras for $N=1,2,3,4$ naturally fit with the study of  Lorentzian Kac-Moody Lie superalgebras associated with rank-three Cartan matrices by Gritsenko and Nikulin\cite{Gritsenko:2002}. Gritsenko and Nikulin show that there exists no BKM Lie superalgebras associated with hyperbolic root systems with Weyl vector of hyperbolic type. The root lattice associated with $A^{(6)}$ is of this type and hence $\mathcal{B}(A^{(6)})$ cannot be a BKM Lie superalgebra. Our long-term goal is construct the  Lie superalgebra $\mathcal{B}(A^{(6)})$, if it is exists, whose WKB superdenominator formula is given by the Siegel modular form for $N=6$. Clearly some additional inputs beyond the Borcherds correction term is needed. We do not solve this problem here but take the first step towards this by decomposing the Siegel modular forms in terms of a $\slhat$ subalgebra of $\mathfrak{g}(A^{(N)})$.

Feingold and Frenkel study  a Lorentzian Kac-Moody Lie algebra associated with a rank three Cartan matrix using a $\slhat$-subalgebra\cite{Feingold1983}.  Inspired by this, we study all the five modular forms in terms of a $\slhat$ subalgebra present in all five examples. This enables us to carry out a systematic decomposition of the Seigel modular forms in terms of characters of $\slhat$ and its Borcherds extension that we denote by $\mathcal{B}_N(\slhat)$ -- this is a sub-algebra of $\mathcal{B}(A^{(N)})$ for $N\neq 6$. \\

\noindent A summary of the main results of this paper is as follows.
\begin{enumerate}
\item We decompose the Umbral Jacobi forms in terms of characters of a $\slhat$ Lie algebra. This leads to vector valued modular forms (vvmfs) with well-defined modular properties that we determine.
\item For $N\leq 4$, we show that these vvmfs arise as solutions to a matrix differential equation proposed by Gannon\cite{Gannon:2013jua}. For $N=6$, we obtain a closed formula for the Fourier coefficients of the vvmf using a different method.
\item We re-express the decomposition of the Umbral Jacobi form in terms of $\mathcal{B}_N(\slhat)$ characters and assign weights using roots of $\mathcal{B}(A^{(N)})$. Using the covariance properties of the Siegel modular forms, we are able identify Weyl orbits for each of the weights that appear in the Lie algebraic decomposition. This provides a preliminary insight into rewriting the Siegel modular forms as sums of Weyl orbits.
\end{enumerate}

The organisation of the paper is as follows. Following the introductory section, in section 2, we introduce the Lorentzian rank three root lattices, their Weyl group as well as the Siegel modular forms that are potential WKB denominator formula for Lie superalgebras. The Siegel modular forms are constructed using a product formula due to Gritsenko-Nikulin with Umbral Jacobi forms as input.  In section 3, we introduce the embedding of $\slhat$ as a subalgebra of the $\mathfrak{g}(A^{(N)})$ Lie algebras. This  naturally leads to the decomposition of the Umbral Jacobi forms in terms of $\slhat$ characters whose coefficients form a vector-valued modular form (vvmf). We determine the first few terms in the Fourier expansion of the vvmf on a computer. In section 4, we obtain more explicit formulae for the vvmf using the approach of Gannon by constructing them as solutions to a modular differential equation. This approach did not work of $N=6$. In section 4.2, we use another method to get an explicit formula for the vvmf. In section 4.3, we rewrite parts of the Siegel modular form in terms of root vectors  that appear in the decomposition of the Umbral Jacobi form. We conclude in section 5 with some remarks. Two appendices provide definitions that are used in the paper.  Appendix A provides the background on the various modular forms, Jacobi forms and Siegel modular forms that appear. Appendix B defines the Weyl-Kac-Borcherds superdenominator formula as well as the supercharacter formula for BKM Lie superalgebras.

 \section{The $\mathcal{B}(A^{(N)})$ BKM Lie superalgebras}
\subsection{Root Lattices, hyperbolic polygons and Weyl groups}

Consider the following two matrices.
\begin{equation}
\alpha_1= \begin{pmatrix} 2 & 1 \\ 1 & 0 \end{pmatrix}\text{ and } \ 
\alpha_2=\begin{pmatrix} 0 & -1 \\ -1 & 0 \end{pmatrix} \ .
\end{equation}
For $m\in \BZ$, define
\begin{equation}
\alpha_{a+2m} = \left(\gamma^{(N)}\right)^m \cdot \alpha_a \cdot \left((\gamma^{(N)})^T\right)^m\text{ for } a=1,2,
\end{equation}
where
\[
\gamma^{(N)}=\left(\begin{matrix}
                1 & -1 \\  N & 1-N
               \end{matrix}  \right)\ .
\]
Using the above definition, observe that
\begin{equation}
\alpha_0 =\begin{pmatrix}
2N-2 & 2N-1 \\ 2N-1 & 2N 
\end{pmatrix} \text{ and } \ 
\alpha_3= \begin{pmatrix} 0 & 1 \\ 1 & 2N \end{pmatrix} .
\end{equation}
Let $\mathbf{X}_N$ denote the ordered sequence of distinct matrices $\alpha_i$ generated in this fashion. These sequences are periodic for $N=1, 2, 3$ and one has
\begin{equation}
\mathbf{X}_N= (\alpha_i) \text{ for }i\in \mathcal{S}_N=\begin{cases}(1,2,3\text{ mod } 3)\ , & N=1 \\
(0,1,2,3\text{ mod } 4)\ , & N=2 \\
(0,1,2,3,4,5\text{ mod } 6)\ , & N=3 \\
\BZ \ , & N=4,6
\end{cases}\ .
\end{equation}
We shall call the elements of $\mathbf{X}_N$ roots in anticipation of the fact that they are indeed the real simple roots of a Lie algebra. 

\begin{figure}[htb] 
\centering
\begin{tikzpicture}[xscale=12,yscale=12]
 \draw[thick,color=blue] (0,0) arc(180:0:1/6) (1/3,0);
 \draw[thick,color=blue] (2/3,0) arc(180:0:1/6) (1,0);
 \draw[thick,color=blue] (0,0) -- (0,0.35);
  \draw[thick,color=blue] (1,0) -- (1,0.35);
  \draw (0,0) -- (1,0);
  \node [label=below:$\tfrac01$] (X) at (0,0) {};  
  \node [label=below:$\tfrac11$]  at (1,0) {};
   \node [label=below:$\tfrac1N$] at (1/3,0) {};
   \node [label=below:$\tfrac{N-1}N$] at (2/3,0) {};
     \node  at (0.97,0.25){$\alpha_1$};
          \node  at (0.03,0.25){$\alpha_2$};
          \node  at (1/4,0.17){$\alpha_3$};
                    \node  at (3/4,0.17){$\alpha_0$};
\end{tikzpicture}
\caption{The four roots $\alpha_0,\alpha_1,\alpha_2,\alpha_3$ represented as semi-circles in the upper half plane.}\label{fourroots}
\end{figure}
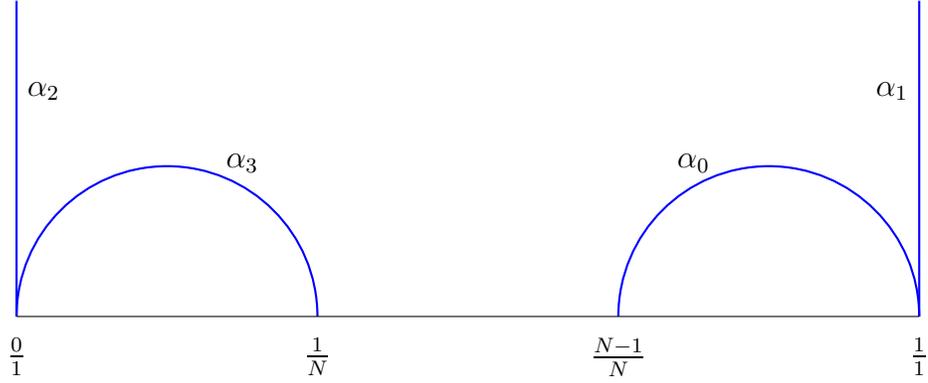

The elements of $\mathbf{X}_N$ are in one-to-one correspondence with edges of a polygon, $\mathcal{M}_N$, in the hyperbolic upper half plane with vertices at rational points in the real line which is the boundary of the upper half plane. Let 
$(\frac{b}{a},\frac{d}{c})$ denote adjacent vertices of the hyperbolic polygon. The root corresponding to the edge connecting these two vertices is given by the map\cite{Cheng:2008fc}
\begin{equation}
\left(\frac{b}{a},\frac{d}{c}\right)  \longrightarrow \alpha=\begin{pmatrix}
2bd & ad+bc \\ ad + bc  & 2ac
\end{pmatrix}\ .
\end{equation}
We illustrate this for some roots in Figure \ref{fourroots}.
For $N=6$ consider the following additional roots. 
\begin{equation}
\left(\tfrac23,\tfrac12\right)\longleftrightarrow\widetilde{\alpha}_1 =\begin{pmatrix} 2 & 5 \\ 5 & 12 \end{pmatrix} \quad ,\quad
\left(\tfrac12,\tfrac23\right)\longleftrightarrow\widetilde{\alpha}_{2} = \begin{pmatrix} 4 & 7 \\ 7 &12  \end{pmatrix}\quad.
\end{equation}
For $m\in \BZ$, define
\begin{equation}
\widetilde{\alpha}_{2m+a} = \left(\gamma^{(6)}\right)^m \cdot \widetilde{\alpha}_a \cdot \left((\gamma^{(6)})^T\right)^m\text{ for } a=1,2.
\end{equation}
Define $\widetilde{\mathbf{X}}_6 = (\widetilde{\alpha}_i)$ for $i\in\BZ$. The two (infinite) sets of  roots combine to give the hyperbolic polygon  $\mathcal{M}_6$ (see Figure \ref{Sixroots}).
\definecolor{darkgreen}{RGB}{10,90,10}
\begin{figure}[htb]
\centering
\begin{tikzpicture}[scale=12]
 \draw[thick,color=blue] (0,0) arc(180:0:1/12) (1/6,0);
 \draw[thick,color=blue] (1/6,0) arc(180:0:1/60) (1/5,0);
 \draw[thick,color=blue] (5/6,0) arc(180:0:1/12) (1,0);
 \draw[thick,color=blue] (4/5,0) arc(180:0:1/60) (5/6,0);
 \draw[thick,color=darkgreen] (1/3,0) arc(180:0:1/12) (1/2,0);
 \draw[thick,color=darkgreen] (1/4,0) arc(180:0:1/24) (1/3,0);
 \draw[thick,color=darkgreen] (1/2,0) arc(180:0:1/12) (2/3,0);
 \draw[thick,color=darkgreen] (2/3,0) arc(180:0:1/24) (3/4,0);
 \draw[thick,color=blue] (0,0) -- (0,0.3);
  \draw[thick,color=blue] (1,0) -- (1,0.3);
  \draw (0,0) -- (1,0);
  \node [label=below:$\tfrac01$] (X) at (0,0) {};  
  \node [label=below:$\tfrac11$]  at (1,0) {};
  \node [label=below:$\tfrac12$] (X) at (1/2,0) {};
   \node [label=below:$\tfrac13$] at (1/3,0) {};
  \node [label=below:$\tfrac16$] at (1/6,0) {};
  \node [label=below:$\tfrac15$] at (1/5,0) {};
   \node [label=below:$\tfrac23$] at (2/3,0) {};
  \node [label=below:$\tfrac14$] at (1/4,0) {};
    \node [label=below:$\tfrac56$] at (5/6,0) {};
  \node [label=below:$\tfrac34$] at (3/4,0) {};
  \node [label=below:$\tfrac45$] at (4/5,0) {};
  \draw (0.7887,0) node{$\bullet$};
  \node [label=below:$\tfrac12(1+\tfrac1{\sqrt3})$] at (0.7887,-0.1) {};
  \draw[-stealth] (0.7887,-0.13) -> (0.7887,-0.015);
 \draw (0.2113,0) node{$\bullet$};
 \node [label=below:$\tfrac12(1-\tfrac1{\sqrt3})$] at (0.2113,-0.1) {};
  \draw[-stealth] (0.2113,-0.135) -> (0.2113,-0.015);
\end{tikzpicture}
 \caption{We show some of the roots in $\mathbf{X}_6$ (in blue)  and $\widetilde{\mathbf{X}}_6$ (in green) which form the edges of the hyperbolic polygon $\mathcal{M}_6$. The two dark circles indicate limit points.}\label{Sixroots}
\end{figure}
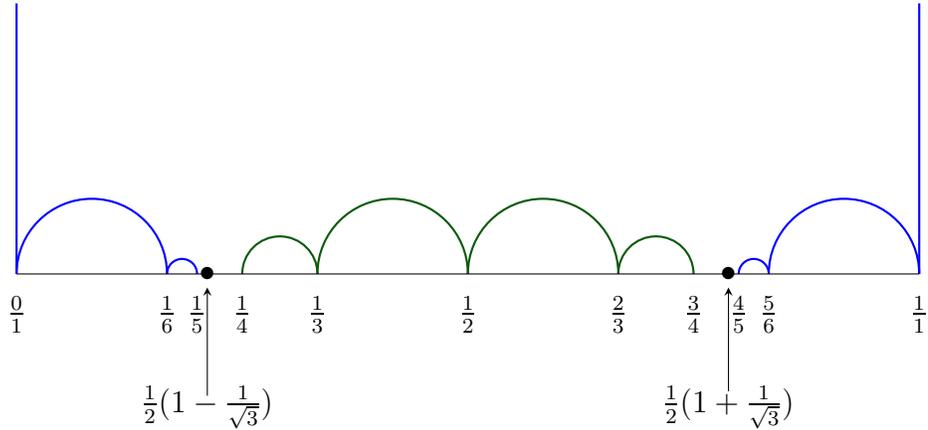

The operation $\hdelta$  acts on the roots as follows:
\begin{equation}\label{deltadef}
\alpha \rightarrow \hdelta \cdot \alpha \cdot \hdelta^{\,T} \quad \textrm{with } \hdelta =\begin{pmatrix} -1 & 1 \\ 0 & 1\end{pmatrix}\ ,
\end{equation}
and $\alpha$ is any root. $\hdelta^{\, 2}$ acts as the identity operator on the roots. It acts on the elements of $\mathbf{X}_N$ as the following involution.
\begin{equation}
 \alpha_{m} \stackrel\hdelta{\longleftrightarrow} \alpha_{3-m}\ .
\end{equation}
The group $\text{Dih}(\mathcal{M}_N)$ generated by $\gamma^{(N)}$ and $\hdelta$ acts as a dihedral symmetry of the hyperbolic polygon $\mathcal{M}_N$.

A real symmetric $2\times 2$ matrix  can be considered as a vector in $\BR^{2,1}$ as follows\cite{Cheng:2008fc}. 
\[
\begin{pmatrix}
x \\ y \\ t
\end{pmatrix}\longleftrightarrow v=\begin{pmatrix}
t + y & x \\ x & t-y
\end{pmatrix}
\]
with norm $\langle v,v \rangle=-2 \det(v)= 2(x^2 + y^2 -t^2)$. The Cartan matrix for the real simple roots is given by the matrix of inner products of the roots of $\mathbf{X}_N$, i.e., $A^{(N)}:=(a_{nm}),$ with $n,m\in \mathcal{S}_N$. A closed formula for the Cartan matrices is given by Eq. \eqref{CartanMatrix}. For $N=6$, where there are additional roots that appear in $\widetilde{X}_6$, the following inner products hold
\begin{equation}\label{innerproducts}
\langle \alpha_n,\alpha_m \rangle= \langle \widetilde{\alpha}_m, \widetilde{\alpha}_n\rangle= \langle \widetilde{\alpha}_m,\alpha_n \rangle+ 12\ .
\end{equation}
  The following relation holds
\begin{equation}
\alpha_m + (N-1)\ \alpha_{m+2} = (N-1)\ \alpha_{m+1} +\alpha_{m+3}\quad \forall m\in \mathcal{S}_N\ .
\end{equation}
This is consistent with the matrices $A^{(N)}$ having rank three.
We will focus on the following four real simple roots $(\alpha_0,\alpha_1,\alpha_2,\alpha_3)$ as we will mostly track the Weyl reflections due to these roots. Using this norm, the matrix of inner products of these roots is
\begin{equation}
A^{(N)}_\text{trun}=\begin{pmatrix}
 ~2 & -2 & (2-4N) & (-2+8N -4N^2)\\
  -2 & ~2 & -2 & (2-4N) \\
    (2-4N) & -2 & ~2 &-2 \\
 (-2+8N -4N^2) & (2-4N) & -2 & ~2
 \end{pmatrix}
\end{equation}
When $N=1$, the first three columns and rows give $A^{(1)}$ as $\alpha_0=\alpha_3$. When $N=2$, $A^{(2)}_\text{trun}=A^{(2)}$ as there are only four real simple roots. In all other cases, we obtain a truncation of the Cartan matrix $A^{(N)}$.

\subsubsection*{The Weyl Group and its extension}

Let $w_i$ denote the elementary Weyl reflection by the simple root $\alpha_i\in \mathbf{X}_N$ and $\widetilde{w}_i$ the Weyl reflection by the simple root $\widetilde{\alpha}_i\in \widetilde{\mathbf{X}}_6$. Let $\beta$ be a root. Then
\[
w_i(\beta) = \beta - 2\,\frac{\langle \alpha_i,\beta\rangle}{\langle \alpha_i,\alpha_i \rangle}\ \alpha_i\ .
\]
Let $W=W(A^{(N)})$ denote the Weyl group generated by elementary Weyl reflections associated with all simple roots in $\mathbf{X}_N$. Let us call the group 
$$
\mathcal{W}:=W(A^{(N)}) \rtimes \text{Dih}(\mathcal{M}_N)\ ,
$$ 
the \textit{extended} Weyl group. The extended Weyl group is generated by $\langle \gamma^{(N)},\hdelta, w_2\rangle$.

\subsection{Root Lattices with Weyl vector}

Consider the vectors contained in $\mathbf{X}_N$ for  $N\leq 4$. For $N=6$, the vectors are given by $\mathbf{X}_N \cup \widetilde{\mathbf{X}}_N$. These generate lattices in $\mathbb{R}^{2,1}$.  For $N\leq 4$, the lattice is given by
\begin{equation}
\mathcal{L}^{(N)} = \left\{\sum_{m \in \mathcal{S}_N} c_m\, \alpha_m ~\big|~ c_m\in \BZ\right\} \ .
\end{equation}
and for $N=6$
\begin{equation}
\mathcal{L}^{(6)} 
= \left\{\sum_{m \in \BZ} c_m\, \alpha_m + \sum_{m \in \BZ} c'_m\, \widetilde{\alpha}_m ~\big|~ c_m, c'_m\in \BZ\right\}\ .
\end{equation}
The $\mathcal{L}^{(N)}$ are all rank-three Lorentzian lattices with lattice Weyl vector 
\begin{equation}
\varrho^{(N)}=\begin{pmatrix} 1/N & 1/2 \\ 1/2 & 1\end{pmatrix}\text{ with } \langle \varrho^{(N)},\varrho^{(N)}\rangle=\frac12 -\frac2N\ .
\end{equation}
The lattice Weyl vector has the following properties:
\begin{enumerate}
\item The norm of $\varrho^{(N)}$ is $\left(\frac{N-4}{2N}\right)$. Thus, the norm is time-like ($<0$) for $N<4$, light-like ($=0$) for $N=4$ and space-like ($>0$) for $N=6$.
\item The inner products of the lattice Weyl vector with real simple roots are:
\[
 \langle \varrho^{(N)}, \alpha_m \rangle=-1\quad  \forall\ \alpha_m \in \mathbf{X}_N
\]
and for $N=6$, additionally one has
\[
 \langle \varrho^{(6)}, \widetilde{\alpha}_m \rangle=+1\quad  \forall\ \widetilde{\alpha}_m \in \widetilde{\mathbf{X}}_6\ .
\]
\item 
The generators of  Dih$(\mathcal{M}_N)$ act on the Weyl vector as follows:
\[
\gamma^{(N)}: \varrho^{(N)}\rightarrow \varrho^{(N)} \ \text{ and }
\hdelta: \varrho^{(N)}\rightarrow \varrho^{(N)}. \  \
\]
Thus Dih$(\mathcal{M}_N)$ preserves the Weyl vector.
\end{enumerate}
The rank-three hyperbolic root lattices $\mathcal{L}^{(N)}$ with lattice Weyl vector $\varrho^{(N)}$  fit with Nikulin's classification of hyperbolic root systems of rank three\cite{Nikulin:2000}. According to that classification, for $N\leq 3$, the lattice Weyl vector is of elliptic type, while for $N=4$ it is of parabolic type and for $N=6$, it is of hyperbolic type. The type is determined by the norm of the lattice Weyl vector.

\subsection{Construction of the Siegel modular forms}

The following theorem enables one to construct Siegel modular forms in the form of a product formula. All the five Siegel modular forms that we study arise in this fashion as we will show. 
\begin{theorem}[Gritsenko-Nikulin\cite{GritsenkoNikulinII}] \label{CGproduct} Let $\psi$ be a nearly holomorphic Jacobi form of weight $0$ and index $t$ with integral Fourier coefficients. 
\[
\psi(\tau,z) = \sum_{n,\ell \in \BZ} c(n,\ell)\ q^n r^\ell\quad,\quad c(n,\ell)\in \BZ \ .
\] Then the product
  \[
  B_\psi(\mathbf{Z}) = q^A r^B s^C  \prod_{\substack{n,\ell,m\in \mathbb{Z}\\ (n,\ell,m)>0}} \Big(1-q^nr^\ell s^{tm}\Big)^{ c(nm,\ell)}\ ,
  \]
 where $(n,l,m)>0$ means that: if $m>0$, then $n\in \BZ$ and $\ell\in \BZ$; if $m=0$, then $n>0$ and $\ell \in \BZ$; if $m=n=0$, then $\ell <0$ and
  \[
  A=\frac1{24} \sum_{ \ell \in \mathbb{Z}}  c(0,\ell) ,\ 
  B=\frac1{2} \sum_{\ell \in \mathbb{Z}_{>0}} \ell \, c(0,\ell),\ 
   C=\frac1{4} \sum_{ \ell \in \mathbb{Z}} \ell^2 \, c(0,\ell)\ ,
  \]
defines a meromorphic Siegel modular form of weight $k=  \frac1{2} \,  c(0,0)$
with respect to $\mathbf{\Gamma}_t^+$ possibly with character. The character is determined by the zeroth Fourier-Jacobi coefficient (i.e., the coefficient of $s^C$) of $B_\psi(\mathbf{Z})$  which is a Jacobi form of weight $k$ and index $C$ of the Jacobi subgroup of $\mathbf{\Gamma}_t^+$.
  \end{theorem}

The five Siegel modular forms of interest are defined as follows. For $N=1,2,3,4,6$, let
\begin{equation}\label{SMFdef}
\Delta_{k(N)}(\mathbf{Z}) := B_{\psi_{0,N}} (\mathbf{Z}) \ ,
\end{equation}
where $\psi_{0,N}(\tau,z)$ are the Umbral Jacobi forms defined in Eq. \eqref{UJFlist}. Using the Fourier expansion given there, we obtain $A=1/2N$, $B=C=1/2$ and $k(N)=(6/N)-1$. The zeroth Fourier-Jacobi coefficient is
\begin{equation}\label{zerothFJ}
\phi_{k(N),1/2}(\tau,z) = \vartheta_1(\tau,z) \eta(\tau)^{2k(N)-1}\ .
\end{equation}
The Jacobi forms transform with character $v_\eta^{12/N}\times v_H$, where $v_\eta$ is the character associated with the Dedekind eta function and $v_H$ is defined in Eq. \eqref{vhdef}. This implies that the Siegel modular forms have the following character\cite{GritsenkoNikulinII}:
\[
v|_{\Gamma_1} = v_\eta^{12/N}\quad,\quad
v|_{H(\BZ)} = v_H \quad,\quad
v|_{[0,0,\kappa/N]} = e^{\pi i \kappa/N}\ .
\]
Further all our Siegel modular forms are symmetric under the operation $V_N:\ q\leftrightarrow s^N$.

\subsubsection*{Properties  of the Siegel modular forms:}

We can translate the action of the Dihedral and Weyl group on the roots to equivalent actions on $\mathbf{Z}=\left(\begin{matrix}
\tau & z \\ z & \tau'
\end{matrix}\right)\in \BH_2$ using the inner product on $\BR^{2,1}$. One has
\[
e^{-\alpha} \longleftrightarrow  e^{2\pi i (\alpha,\mathbf{Z}')} = 
e^{2\pi i \text{Tr}(\alpha\,\mathbf{Z})}\ .
\]
where $\mathbf{Z'}=\det(\mathbf{Z})\, \mathbf{Z}^{-1}=\left(\begin{matrix}
\tau' & -z \\ -z & \tau
\end{matrix}\right)$.

All transformations are realised as elements of $\Gamma_N$ and all that one has to do is to compute the character for those elements. A practical method is to directly compare terms in the Fourier expansion of the Siegel form that are related by the generator.
\begin{equation}\label{covariance}
\begin{split}
\Delta_{k(N)}(\gamma^{(N)}\circ \mathbf{Z}) & =+ \Delta_{k(N)}(\mathbf{Z})\ , \\
\Delta_{k(N)}(\hdelta\circ \mathbf{Z}) & = + \Delta_{k(N)}(\mathbf{Z})\ , \\
\Delta_{k(N)}(w_2\circ \mathbf{Z}) & = -\Delta_{k(N)}(\mathbf{Z})\ .
\end{split}
\end{equation}
where 
$$
\gamma^{(N)}\circ \mathbf{Z}:= ((\gamma^{(N)})^T)^{-1}\cdot \mathbf{Z} \cdot (\gamma^{(N)})^{-1} \text{ and } \hdelta\circ \mathbf{Z}:= (\hdelta^{\, T})^{\, -1} \cdot \mathbf{Z}\cdot  (\hdelta)^{-1}\ ,
$$
and $w_2\circ \mathbf{Z}$ acts on $\BH_2$ as $z\rightarrow -z$ and leaves $(\tau,\tau')$ invariant.
These properties imply that the Siegel modular forms transform covariantly under the extended Weyl group. The proof of these properties is given, for instance, in \cite{GritsenkoNikulinII,Govindarajan:2010fu, Govindarajan:2019ezd}.

\section{Deconstructing the Denominator Formula}

It is known that for $N=1,2,3,4$, the Siegel modular forms $\Delta_{k(N)}(\mathbf{Z})$ are the modular forms associated with the WKB superdenominator formulae for the Borcherds extensions $\mathcal{B}(A^{(N)})$ of $\mathfrak{g}(A^{(N)})$\cite{GritsenkoNikulinI,GritsenkoNikulinII,Govindarajan:2010fu}. Eq. \eqref{covariance} shows that the Siegel modular forms transform as expected from a WKB superdenominator formula. Further, it is easy to see from explicit formulae that one has
\begin{equation}
\Delta_{k(N)}(\mathbf{Z}) = \sum_{w\in W} \det(w)\ w\left( e^{-\varrho^{(N)}}\right) + \cdots 
\end{equation}
The simple roots $(\alpha_1,\alpha_2)$ generate a $\slhat$ sub-algebra. We study this family of Siegel modular forms in terms of this  $\slhat$ sub-algebra as well as a Borcherds extension of it, that we call $\mathcal{B}_N(\slhat)$, obtained by adding imaginary simple roots of the form $n\delta=n(\alpha_1+\alpha_2)$ for $n\in \BZ_{>0}$.

\subsection{Embedding $\widehat{sl(2)}$ in the Lie algebra $\mathfrak{g}(A^{(N)})$}

\subsubsection{Defining $\slhat$}\label{sec:sl2hatdef}

 Let $(e,h,f)$ be the generators of the $sl(2)$ Lie algebra. The non-zero Lie brackets are:
 \[
 [e,f]=h\quad,\quad [h,e]=2e \quad,\quad [h,f]=-2f\ ,
 \]
 and the (normalised) Killing form is $\langle e,f\rangle=1$ and $\langle h,h\rangle=2$.
 
 The affine Lie algebra $\widehat{sl(2)}$ is defined by
 \[
 \widehat{sl(2)} = sl(2)\otimes\mathbb{C}[t,t^{-1}]\oplus \mathbb{C}\,\hat{k} \oplus \mathbb{C}\, d\ ,
 \]
 where $\hat{k}$ is the central extension and $d=-t d/dt$. The Lie algebra (with $x\in sl(2)$)
 \begin{align*}
 [x\otimes t^n,y\otimes t^m] &= [x,y]\otimes t^{n+m} + n\, \langle x,y\rangle \ \hat{k}\, \delta_{n+m,0} \\
 [d, x\otimes t^{n}] &= -n \, x\otimes t^{n}
 \end{align*}
 The Cartan sub-algebra is $(h\otimes 1, \hat{k},d)$ with inner product such that $\langle h\otimes 1, h\otimes 1\rangle=2$ and $\langle \hat{k},d\rangle =-1$.

We would like to embed $\widehat{sl(2)}$ into the Kac-Moody Lie algebra $\mathfrak{g}(A^{(N)})$ with symmetric Cartan matrix given by $A^{(N)}$. Consider the Chevalley generators $\{e_i,f_i,h_i~|~i=1,2,3\}$ corresponding to the real simple roots $\alpha_1, \alpha_2, \alpha_3$ of $\mathfrak{g}(A^{(N)})$. These satisfy
\[
 [h_i, h_j]=0\ ,\  [h_i,e_j] = a_{ji} \ e_i\ ,\   [e_i,f_i] = h_i\ ,
\]
where $(a_{ij})$ is the Cartan matrix for these three roots
 \[
\begin{pmatrix}
 2 & -2 & (2-4N)  \\
 -2 & 2 & -2  \\
 (2-4N) & -2 & 2  
 \end{pmatrix}\ .
\]
The Lie subalgebra of $\mathfrak{g}(A^{(N)})$ generated by $e_1, f_1, e_2, f_2, h_1, h_2$ and $h_3$ is isomorphic to $\widehat{sl(2)}$. Following Feingold and Frenkel\cite{Feingold1983}, we make the identification
 \[
 e\otimes 1 =e_2\ ,\  f\otimes 1=f_2\ ,\  f \otimes t =e_1\ , \  e\otimes t^{-1} =f_1 \ ,
 \]
 For the Cartan subalgebra of $\widehat{sl(2)}$, using the above identification, we obtain
 \[
 h_1 = [e_1,f_1]=-h\otimes 1 +\hat{k}\ , \ h_2 = [e_2,f_2]= h \otimes 1 \ ,\ h_3= - h\otimes 1  +4N\, d\ .
 \]
The inverse is
\[
h\otimes 1 = h_2\ , \  \hat{k} = h_1 + h_2 \ , \ d =  \frac{1}{4N} (h_2 + h_3) \ .
\]

\subsubsection{Defining $\mathcal{B}_N(\widehat{sl(2)})$}

Let $\delta=\alpha_1+\alpha_2$. Consider the Borcherds extension of $\slhat$ obtained by the addition of imaginary simple roots, $(\delta,2\delta,\ldots)$  with zero norm, each with multiplicity $(2k(N)-1)=(12-3N)/N$. Let us call this extension $\mathcal{B}_N(\widehat{sl(2)})$. Note that $\mathcal{B}_4(\widehat{sl(2)})=\slhat$ i.e., no imaginary simple roots are added when $N=4$.

\subsection{Characters}
As before, let $W=W(A^{(N)})$ denote the Weyl group of $\mathfrak{g}(A^{(N)})$; this is generated by the  elementary reflections $w_i$ for all simple roots  $i \in \mathbf{X}_N$. 
Let $\widehat{W}$ be the subgroup generated by the reflections $w_1, w_2$, corresponding to the simple roots $\alpha_1$ and $\alpha_2$. This is isomorphic to the Weyl group of $\widehat{sl(2)}$, with
\[
\widehat{W} : = \{(w_1\cdot w_2)^k, w_2\cdot (w_1\cdot w_2)^k ~|~ k\in \mathbb{Z}\}
\]
We have
\[
W = 
\widehat{W}\cup (\widehat{W}\cdot w_0) \cup (\widehat{W}\cdot w_3) \cup (w_0 \cdot \widehat{W}) \cup (w_3 \cdot \widehat{W})\cup \cdots
\]
We will be focusing mostly on the first three terms.
The numerator of the Weyl character formula for a highest weight state with weight $\tilde{\Lambda}$ of $\mathcal{B}_N(\slhat)$  takes the form
\begin{align}\label{numerator}
\textrm{num}(\tilde\Lambda) &= \sum_{w\in W} \det(w) w\left(e^{-\varrho^{(N)} -\Lambda}\,T_{\tilde\Lambda}\right) \nonumber \\
&= \sum_{w\in \widehat{W}} \det(w) w\left(e^{-\varrho^{(N)} -\tilde{\Lambda}}\,T_{\tilde\Lambda}\right)
-\sum_{w\in \widehat{W}} \det(w) w\left(e^{-w_3(\varrho^{(N)} +\tilde{\Lambda)}}\,T_{\tilde{\Lambda}}\right) + \cdots
\end{align}
Note that  num$(0)$ is the denominator formula. We have  included imaginary simple roots (the Borcherds correction factor defined in Sec. \ref{sec:supercharacter}) in the above formula as $T_{\tilde\Lambda}$.\\

\noindent Let $\delta=(\alpha_1+\alpha_2)$. A straightforward computation gives the following formulae.
\begin{align*}
(w_1\cdot w_2) (\varrho^{(N)}) &= \varrho^{(N)} + 3 \delta -2\alpha_2\\
(w_1\cdot w_2) (\delta) &= \delta\ , \\
(w_1\cdot w_2) (\alpha_2) &= -2 \delta + \alpha_2\\
(w_1\cdot w_2) (\alpha_3) &= \alpha_3 + (4N+2) \delta - 4N \alpha_2 \\
(w_1\cdot w_2) (\alpha_0) &= \alpha_0 + (4N-1) \delta - \alpha_2
\end{align*}
Using this computation, we can show that for $\tilde\Lambda =a \delta + b \alpha_2 + c \alpha_3 $, one has (with $m=(4Nc+2)$ and $\ell=(2c-2b)$)
\begin{align*}
(w_1\cdot w_2)^k \Big[ \varrho^{(N)} + \tilde\Lambda \Big]-(\varrho + \tilde\Lambda) &=
\Big[ \big( mk^2+(\ell+1)k \big) \delta + (-k m )\alpha_2 \Big] \\
w_2\cdot(w_1\cdot w_2)^k \Big[ \varrho^{(N)} + \tilde\Lambda \Big]-(\varrho + \tilde\Lambda) &=
\Big[ \big( mk^2+(\ell+1)k \big) \delta + (k m+\ell+1 )\alpha_2 \Big] 
\end{align*}
 After making the following identifications: $$e^{-\delta}\sim q=\exp(2\pi i\tau), \ e^{-\alpha_2}\sim r^{-1}=\exp{(-2\pi i z)}\text{ and }e^{-\alpha_3}\sim s^Nr\ ,$$
 we obtain the master formula 
\begin{equation}\label{masterformula}
\frac{\sum_{w\in \widehat{W}} \det(w) w\left(e^{-\varrho^{(N)}-\tilde\Lambda}\right)}{\left(e^{-\varrho^{(N)}-\tilde\Lambda}\right)}  = \,q^{-\frac{(\ell+1))^2}{4m}}r^{-\frac{(\ell+1)}2)}
\left(\theta_{m,\ell+1}(\tau,z)-\theta_{m,-\ell-1}(\tau,z)\right)
\end{equation}
where
\begin{equation}
\theta_{m,a}(\tau,z) := \sum_{k\in \mathbb{Z}} q^{m(k +\frac{a}{2m})^2} r^{m(k +\frac{a}{2m})}\ .
\end{equation}

\noindent \textbf{The denominator formula for $A^{(N)}$:}\\

It is known that for $N=1,2,3,4$, the genus two Siegel modular form $\Delta_{k(N)}(\mathbf{Z})$ arises as the WKB denominator formula for an extension of the Kac-Moody Lie algebra $\mathfrak{g}\big(A^{(N)}\big)$ by the addition of imaginary simple roots. For $N=1,2,3,4,6$, it is known that the modular forms admit the following expansion\cite{Govindarajan:2010fu, Govindarajan:2019ezd}:
\begin{equation}
\Delta_{k(N)}(\mathbf{Z}) = s^{1/2}\phi_{k(N),1/2}(\tau,z) \Big[1 - s^N\ \psi_{0,N}(\tau,z) + O(s^{2N})  \Big]\ ,
\end{equation}
where $\phi_{k(N),1/2}(\tau,z)$ is defined in Eq. \eqref{zerothFJ}
and $\psi_{0,N}(\tau,z)$ is an Umbral Jacobi form defined in Eq. \eqref{UJFlist}(see \cite{Cheng:2012tq} for its connection to Umbral Moonshine). It is important to observe that terms in the square brackets, for each power of $s$, are invariant under the Weyl group $\widehat{W}$ of $\slhat$. As we will show next, we can expand each of the terms in terms of characters of $\slhat$ and of $\mathcal{B}_N(\widehat{sl(2)})$. One should keep in mind that the two characters are different.


We interpret $\phi_{k(N),1/2}(\tau,z)$ as the WKB denominator formula for the Borcherds extension  $\mathcal{B}_N(\widehat{sl(2)})$.
Then, using the following result that follows from Eq. \eqref{masterformula}
\[
\sum_{w\in \widetilde{W}} \det(w) w\left(e^{-\varrho^{(N)}-a \delta}\right)  = q^a\, [q^{1/N} r^{1/2} s^{1/2}]\ q^{-1/8}r^{-1/2}
\left(\theta_{2,-1}(\tau,z)-\theta_{2,1}(\tau,z)\right)\ ,
\]
we see that 
\begin{equation}
s^{1/2}\,\phi_{k(N),1/2}(\tau,z) = \sum_{w\in \widehat{W}} \det(w) w\left(e^{-\varrho^{(N)}}\ \widetilde{T}\right)
\end{equation}
with the Borcherds correction defined in Eq. \eqref{eq:sumside} given by $\widetilde{T} = \prod_{p=1}^\infty (1-e^{-p\delta})^{2k(N)-1}$.
 The Weyl character formula for the weight vector  $\widetilde{\Lambda} = a \delta + b \alpha_2 + c \alpha_3$ of  $\mathcal{B}_N(\widehat{sl(2)})$ with $\langle \widetilde{\Lambda},\delta\rangle< 0$ is obtained by applying the formula for the supercharacter given in Eq. \eqref{WeylCharacterFormula}.\footnote{The condition $\langle \widetilde{\Lambda},\delta\rangle< 0$ ensures that there is no Borcherds correction term in the numerator of the character formula.} We obtain
\begin{equation}
\widetilde{\chi}_{\widetilde{\Lambda}}(\tau,z) = \frac{q^{a -\frac{(\lambda+1))^2}{4(\hat{k}+2)}+\frac18}}{\varphi(\tau)^{2k(N)-1}} \ \chi_{\hat{k},\lambda}(\tau,z) \ ,
\end{equation}
where $\varphi(\tau)=\prod_{m=1}^\infty (1-q^m)$, $\hat{k}=4Nc$, $\lambda=(2c-2b)$ and  the normalized $\widehat{sl(2)}$ character $\chi_{\hat{k},\lambda}(\tau,z)$ is defined by
\begin{align}
\chi_{\hat{k},\lambda}(\tau,z)&=\frac{\theta_{\hat{k}+2,\lambda+1}(\tau,z)-\theta_{\hat{k}+2,-\lambda-1}(\tau,z)}{\theta_{2,1}(\tau,z)-\theta_{2,-1}(\tau,z)} \text{ for } \hat{k},\lambda \in \mathbb{Z}_{\geq 0} \text{ and } \lambda\leq \hat{k}\ .
\end{align}
\textbf{Remark:} The $\BZ_2$ outer automorphism of the $\widehat{sl(2)}$ Lie algebra corresponding to $\alpha_1\leftrightarrow\alpha_2$ is represented by the generator $\hdelta$ (see Eq. \eqref{deltadef}) of the Dihedral group, Dih$(\mathcal{M}_N)$. This exchanges the characters $\chi_{4N,j}$ and $\chi_{4N,4N-j}$.  \\

\noindent \textbf{Some examples of interest}\\

Applying the master formula Eq. \eqref{masterformula} to the real simple roots $\widetilde{\Lambda}=\alpha_0$ and $\widetilde{\Lambda}=\alpha_3$ as well as 
the imaginary simple roots (with zero norm and multiplicity $[2k(N)-1]$) $\widetilde{\Lambda}=(\alpha_0+\alpha_1)$ and $\widetilde{\Lambda}=(\alpha_2+\alpha_3)$, we obtain (after dropping a pre-factor of $s^N$ that is present in all terms)
\begin{align*}
\widetilde{\chi}_{\alpha_3} &= \frac{q^{(N-4)/(8N+4)}(\theta_{4N+2,3}-\theta_{4N+2,-3})}{(\theta_{2,1}-\theta_{2,-1})\, \varphi(\tau)^{2k(N)-1}} =\frac{q^{(N-4)/(8N+4)}\chi_{4N,2}}{\varphi(\tau)^{2k(N)-1}} \\
\widetilde{\chi}_{\alpha_0} &= \frac{q^{(N-4)/(8N+4)}(\theta_{4N+2,4N-3}-\theta_{4N+2,-4N+3})}{(\theta_{2,1}-\theta_{2,-1})\, \varphi(\tau)^{2k(N)-1}} =\frac{q^{(N-4)/(8N+4)}\chi_{4N,4N-2}}{\varphi(\tau)^{2k(N)-1}} \\
\widetilde{\chi}_{\alpha_0+\alpha_1} &= \frac{q^{N/(8N+4)}(\theta_{4N+2,4N+1}-\theta_{4N+2,-4N-1})}{(\theta_{2,1}-\theta_{2,-1})\, \varphi(\tau)^{2k(N)-1}} =\frac{q^{N/(8N+4)}\chi_{4N,4N}}{\varphi(\tau)^{2k(N)-1}}\\
\widetilde{\chi}_{\alpha_2+\alpha_3} &= \frac{q^{N/(8N+4)}(\theta_{4N,1}-\theta_{4N,-1})}{(\theta_{2,1}-\theta_{2,-1})\,\varphi(\tau)^{2k(N)-1}} =\frac{q^{N/(8N+4)}\chi_{4N,0}}{\varphi(\tau)^{2k(N)-1}}
\end{align*}
In the sequel, we will refer to $\widetilde{\chi}_{\alpha_3}$ as $\widetilde{\chi}_{4N,2}$ and so on. This extends the labels that we use for $\widehat{sl(2)}$ characters to characters of  $\mathcal{B}_N(\widehat{sl(2)})$.
\subsection{Lie algebra decompositions of Umbral Jacobi Forms}\label{decomposition}

For the cases of interest, we wish to decompose the Umbral Jacobi Forms in terms of characters of $\widehat{sl(2)}$ and $\mathcal{B}_N(\widehat{sl(2)})$. The decomposition takes the form
\begin{align}
\psi_{0,N}(\tau,z) &= \sum_{j=1}^{2N} g_j(\tau)\, \chi_{4N,2N+2-2j}(\tau,z)\ , \\
&= \sum_{j=1}^{2N} f_j(\tau)\, \widetilde{\chi}_{4N,2N+2-2j}(\tau,z)\ ,
\end{align}
Further, one observes that $g_j(\tau)=g_{2N-j}(\tau)$. This follows from the $\BZ_2$ outer automorphism under which $\alpha_1\leftrightarrow\alpha_2$ and $\alpha_0\leftrightarrow \alpha_3$. Thus one has $(N+1)$ independent functions that we organize into a vector $\mathbf{g}:=(g_1,g_2,\ldots,g_{N+1})^T$. Using the modular properties of the normalized characters and the Umbral Jacobi form, we can show that $\mathbf{g}(\tau)$ is a weight zero vector-valued modular form (vvmf) with the following modular properties:
\begin{align}
\mathbf{g}(\tau+1) &=T \cdot \mathbf{g}(\tau)\quad ,\quad
\mathbf{g} \left(-1/\tau\right)=S \cdot
\mathbf{g}(\tau)\ .
\end{align}
The above transformations define the matrices $T$ and $S$.

\subsubsection{$N=1$}

In this case, $\alpha_0=\alpha_3$. Using the above results, we obtain the following expansion:
\begin{equation}
\psi_{0,1}(\tau,z) = f_1(\tau)\ \widetilde{\chi}_{\alpha_3}(\tau,z) + f_2(\tau) \Big[ \widetilde{\chi}_{\alpha_1+\alpha_3}(\tau,z) + \widetilde{\chi}_{\alpha_2+\alpha_3}(\tau,z)\Big]
\end{equation}
with $f_1(\tau) = 1 -92 q + 54 q^2 -85 q^3 + \cdots$ and
$f_2(\tau) = 9 (1 +10q + 11q^2 -73q^3 +\cdots)$. The terms that appear at order $q$ and higher in $f_1$ and $f_2$ are due to imaginary simple roots with negative norm. The expansion in terms of $\slhat$ characters is given by
\begin{equation}
\psi_{0,1}(\tau,z) = g_1(\tau)\ \chi_{4,2}(\tau,z) + g_2(\tau) \Big[ \chi_{4,0}(\tau,z) + \chi_{4,4}(\tau,z)\Big]
\end{equation}One has
\begin{equation}
\mathbf{g}(\tau) = \begin{pmatrix}
q^{-\frac{1}{4}}\left(1-84 q-729 q^2-4366 q^3-19935 q^4-77274 q^5-264610 q^6 + \cdots \right)\\
9\ q^{-\frac{11}{12}}\left(q+19 q^2+155 q^3+821 q^4+3541 q^5+13082 q^6+\cdots \right) 
\end{pmatrix}
\end{equation}
For the above rank-two vvmf, we obtain the following $T$ and $S$ matrices using the known transformations of the normalized characters given in Eq. \eqref{chitransform}
\begin{align}
 T=\begin{bmatrix}e^{-\frac{i\pi}{2}}&0\\0&e^{\frac{i\pi}{6}}\end{bmatrix}\quad,\quad S=\frac{1}{\sqrt{3}}\begin{bmatrix} 
-1 & 2\\ 
1 & 1 
\end{bmatrix}\ .
\end{align}

\subsubsection{$N=2$}

The Umbral Jacobi form at lambency $3$ has the following decomposition in terms of $\widehat{sl(2)}$ characters:
\begin{align*}
\psi_{0,2}(\tau,z)&= g_1(\tau)\chi_{8,4}(\tau,z)+g_2(\tau)\left[ \chi_{8,2}(\tau,z) +\chi_{8,6}(\tau,z)\right]+g_3(\tau)\left[ \chi_{8,0}(\tau,z) +\chi_{8,8}(\tau,z)\right]   \\
&= f_1(\tau) (\tilde{\chi}_{\alpha_1+\alpha_3}+\tilde{\chi}_{\alpha_0+\alpha_2}) + f_2(\tau) (\tilde{\chi}_{\alpha_3}+\tilde{\chi}_{\alpha_0})+f_3(\tau) (\tilde{\chi}_{\alpha_2+\alpha_3}+\tilde{\chi}_{\alpha_0+\alpha_1})\ ,
\end{align*}
where
\[
f_1(\tau) = \frac{g_1(\tau)\varphi(\tau)^3}{2q^{1/2}}\quad,\quad
f_2(\tau) = \frac{g_2(\tau)\varphi(\tau)^3}{q^{-1/10}}\quad,\quad
f_3(\tau) = \frac{g_3(\tau)\varphi(\tau)^3}{q^{1/10}}\quad.
\]
Note that since the Cartan matrix $A^{(2)}$ has rank three, one has the identity $\alpha_0+\alpha_2=\alpha_1+\alpha_3$. Using the symmetry generated by $\hdelta$, we write the coefficient of $f_1$ as $(\tilde{\chi}_{\alpha_1+\alpha_3}+\tilde{\chi}_{\alpha_0+\alpha_2})$ anticipating that these two terms will be distinct if we make the Cartan matrix invertible. The weight vector $(\alpha_0+\alpha_2)$ is associated with an imaginary simple root of negative norm.

The vvmf $\mathbf{g}(\tau)$ has rank three and the first few terms in the Fourier expansion are:
\begin{equation}
\mathbf{g}(\tau) = \begin{pmatrix}
-10 q^{-\frac{1}{2}}\left(q+3 q^2+9 q^3+22 q^4+51 q^5+105 q^6+\cdots\right) \\
q^{-\frac{1}{10}}\left(1+3 q+18 q^2+38 q^3+99 q^4+207 q^5+438 q^6+\cdots \right)\\
q^{-\frac{9}{10}}\left(3q+16 q^2+48 q^3+129 q^4+294 q^5+642 q^6+\cdots \right)
\end{pmatrix} \ .
\end{equation}
For the above rank-three vvmf, using Eq. \eqref{chitransform} we obtain the following $T$ and $S$ matrices:
\begin{align}
 T=\begin{bmatrix}e^{i \pi}&0&0\\0&e^{-\frac{i\pi}{5}}&0\\0&0& e^{\frac{i\pi}{5}}\end{bmatrix}\quad,\quad S=\frac{1}{\sqrt{5}}\begin{bmatrix}
  ~~1 & -2 &  2\\ 
-1 & \frac{\sqrt{5}-1}{2}& \frac{\sqrt{5}+1}{2}\\ 
 ~~1 & \frac{\sqrt{5}+1}{2}& \frac{\sqrt{5}-1}{2}
\end{bmatrix}\ .
\end{align}

\subsubsection{$N=3$}

The Umbral Jacobi Form at lambency $4$ has the following decomposition in terms of $\widehat{sl(2)}$ characters:
\begin{multline}
\psi_{0,3}(\tau,z)= g_1(\tau)\chi_{12,6}(\tau,z)+g_2(\tau)\left[ \chi_{12,4}(\tau,z) +\chi_{12,8}(\tau,z)\right]\\ +g_3(\tau)\left[ \chi_{12,2}(\tau,z) +\chi_{12,10}(\tau,z)\right]
 +g_4(\tau)\left[ \chi_{12,12}(\tau,z) +\chi_{12,0}(\tau,z)\right].   
\end{multline}

Define
\[
f_1 = \frac{g_1(\tau)\varphi(\tau)}{2q^{1/4}}\quad,\quad
f_2 = \frac{g_2(\tau)\varphi(\tau)}{q^{19/28}}\quad,\quad
f_3 = \frac{g_2(\tau)\varphi(\tau)}{q^{-1/28}}\quad,\quad
f_4 = \frac{g_3(\tau)\varphi(\tau)}{q^{3/28}}\quad.
\]
Then the decomposition in terms of $\mathcal{B}_3(\slhat)$ characters is given by the following weights.
\begin{center}
\begin{tabular}{c|c|c|c|c|}
Label & 1 & 2 & 3 & 4  \\ \hline 
\multirow{2}{*}{Weights} &   $\alpha_1+5\alpha_2+\alpha_3$ & $\alpha_1+\alpha_3$ & $\alpha_3$& $\alpha_2+\alpha_3$ \\ ]
 &  $\alpha_0+5\alpha_1+\alpha_2$ & $\alpha_0+\alpha_2$ & $\alpha_0$& $\alpha_0+\alpha_1$ \\ \hline
 Norm &$-6$ &$-16$ &2 & 0 \\
\end{tabular}
\end{center}
 Labels 1 and 2 are associated with imaginary simple roots. The real simple roots $\alpha_0$ and $\alpha_1$ appear with label 3 and the zero-norm imaginary roots $(\alpha_0+\alpha_1)$ and $(\alpha_2+\alpha_3)$ appear with label 4.

The vvmf $\mathbf{g}(\tau)$ has rank four and the first few terms in the Fourier expansion are:
\begin{equation}
\mathbf{g}(\tau) = \begin{pmatrix}
-2 q^{-\frac{3}{4}}\left(q+q^2+2 q^3+3 q^4+5 q^5+7 q^6+11 q^7+\cdots\right) \\
-q^{-\frac9{28}}\left(q+q^3+2 q^4+3 q^5+3 q^6+6 q^7+\cdots\right)\\
q^{-\frac{1}{28}}\left(1+2 q+3 q^2+5 q^3+8 q^4+11 q^5+17 q^6+\cdots\right) \\
q^{-\frac{25}{28}}\left(q+q^2+3 q^3+3 q^4+6 q^5+8 q^6+13 q^7+\cdots\right)
\end{pmatrix}\ .
\end{equation}
For the above rank-three vvmf, we obtain the following $T$ and $S$ matrices:
\begin{align}
 T&=\text{Diag}\Big(e^{\frac{i \pi}{2}}, e^{i\pi \left(\frac{19}{14}\right)},     e^{-\frac{i \pi}{14}}, e^{i\pi \left(\frac{3}{14}\right)}\Big) \quad, \nonumber\\
  S&=\frac{1}{\sqrt{7}} \begin{bmatrix}
  -1 & 2 & - 2& 2\\ 
 ~~1 & -2\sin(\frac{3\pi}{14})& -2\sin(\frac{\pi}{14}) & 2\cos(\frac{\pi}{7})\\
  -1 &-2\sin(\frac{\pi}{14}) & 2\cos(\frac{\pi}{7}) &2\sin(\frac{3\pi}{14}) \\
 ~~1 & 2\cos(\frac{\pi}{7})& 2\sin(\frac{3\pi}{14}) &2\sin(\frac{\pi}{14}) 
\end{bmatrix} \ .
\end{align}

\subsubsection{$N=4$}

The Umbral Jacobi form at lambency $5$ has the following decomposition in terms of $\widehat{sl(2)}$ characters:
\begin{equation}
\psi_{0,4}(\tau,z)= -\chi_{16,8}(\tau,z)+ (\chi_{16,2}(\tau,z) +\chi_{16,14}(\tau,z)) .
\end{equation}
The character $\chi_{16,8}(\tau,z)$ is identified with a bosonic simple root corresponding to $qr^{-4}s^4$ of zero norm. Recall that $\mathcal{B}_4(\slhat)=\slhat$ as there were no imaginary simple roots added to $\slhat$. So this is the first appearance of an imaginary simple root.
The above linear combination is invariant under $T$ as it only contains integral powers of $q$. Under the $S$ transform, it  maps itself up to the phase associated with the index. Hence, there is only one overall constant which we fix by using explicit formulae. 

We also see that the characters $\chi_{16,0}(\tau,z)$ and $\chi_{16,16}(\tau,z)$ do \textit{not} appear in the above expansion.
This is consistent with the fact that there were no imaginary simple roots added to $\slhat$ in this case. So we do not expect imaginary simple roots $(\alpha_i+\alpha_{i+1})$ for $i=0,1,2$ to be present. This is borne out in the $\slhat$ decomposition of $\psi_{0,4}$. However, we find a new imaginary simple root of zero norm appearing with character $\chi_{16,8}(\tau,z)$. It is represented by
the matrix
\[
\begin{pmatrix}
2 & -4 \\ -4 & 8
\end{pmatrix}  \longleftrightarrow   (\alpha_1+6\alpha_2+\alpha_3) \ .
\]
The imaginary root $(\alpha_2 + 6 \alpha_1 + \alpha_0)$ with zero norm also appears in the expansion of $\chi_{16,8}(\tau,z)$.

\subsubsection{$N=6$}

As mentioned earlier, we do not have a BKM Lie superalgebra associated with this Cartan matrix. We do know that the Siegel modular form transforms suitably under the extended Weyl group. In particular, one can see that the Umbral Jacobi form is invariant under $\widehat{W}$ and $\hdelta$. The $\slhat$ characters may be viewed as providing a basis for expanding the Jacobi form. 
The Umbral Jacobi form at lambency $7$ has the following decomposition in terms of $\widehat{sl(2)}$ characters. 
\begin{multline}
\psi_{0,6}(\tau,z)= g_1(\tau)\chi_{24,12}(\tau,z)+g_2(\tau)\left[ \chi_{24,10}(\tau,z) +\chi_{24,14}(\tau,z)\right]\\ 
\qquad +g_3(\tau)\left[ \chi_{24,8}(\tau,z) +\chi_{24,16}(\tau,z)\right]+g_4(\tau)\left[ \chi_{24,6}(\tau,z) +\chi_{24,18}(\tau,z)\right]\\ 
\qquad + g_5(\tau)\left[ \chi_{24,4}(\tau,z) +\chi_{24,20}(\tau,z)\right]+g_6(\tau)\left[ \chi_{24,2}(\tau,z) +\chi_{24,22}(\tau,z)\right]\\
+g_7(\tau)\left[ \chi_{24,0}(\tau,z) +\chi_{24,24}(\tau,z)\right]  \label{psi06}
\end{multline} 
We rewrite the above formula in terms of $\mathcal{B}_6(\slhat)$ characters using the following definition.
\begin{multline}
f_1 = \frac{g_1(\tau)}{q^{1/2}\varphi(\tau)}\ ,\ 
f_2 = \frac{g_2(\tau)}{q^{-1/26}\varphi(\tau)}\ ,\ 
f_3 = \frac{g_2(\tau)}{q^{9/26}\varphi(\tau)}\ ,\ 
f_4 = \frac{g_3(\tau)}{q^{43/26}\varphi(\tau)}\ ,\\
f_5 = \frac{g_2(\tau)}{q^{23/26}\varphi(\tau)}\ ,\ 
f_6 = \frac{g_2(\tau)}{q^{1/26}\varphi(\tau)}\ ,\ 
f_7 = \frac{g_3(\tau)}{q^{3/26}\eta(\tau)}\ .
\end{multline}

In the table below, we give the roots associated with the weight for each of the seven characters. 
\begin{center}
\begin{tabular}{c|c|c|c|c|c|c|c|c|c}
Label & 1 & 2 & 3 & 4 & 5 & 6 & 7 \\ \hline 
\multirow{2}{*}{Weights} & $~~\widetilde{\alpha}_1 + \alpha_1$ & $\widetilde{\alpha}_1$ & $~~\widetilde{\alpha}_1 + \alpha_2$ & $2\alpha_1+\alpha_3$ & $\alpha_1+\alpha_3$ & $\alpha_3$& $\alpha_2+\alpha_3$ \\
 & $=\widetilde{\alpha}_2 + \alpha_2$ &  $\widetilde{\alpha}_2$ & $~~\widetilde{\alpha}_2 + \alpha_1$ & $\alpha_0+2\alpha_2$ & $\alpha_0+\alpha_2$ & $\alpha_0$& $\alpha_0+\alpha_1$ \\ \hline
 Norm & $-24$ & 2 & $-16$  &$-78$& $-40$ & 2 & 0 \\
\end{tabular}
\end{center}
We assume that there is indeed a Lie superalgebra $\mathcal{B}(A^{(6)})$ and make the following statements based on that assumption.  The real simple roots $\widetilde{\alpha}_1$ and $\widetilde{\alpha}_2$ are associated with label 2. Labels 1, 3, 4 and 5 correspond to imaginary simple roots with negative norm. 

The vvmf $\mathbf{g}(\tau)$ has rank seven and the first few terms in the Fourier expansion are:
\begin{equation}
\mathbf{g}(\tau) = \begin{pmatrix}
2 q^{-\frac{1}{2}}\left(q+q^2+2 q^3+3 q^4+5 q^5+7 q^6+11 q^7+\cdots\right) \\
-q^{-\frac{1}{26}}\left(1+q+3 q^2+4 q^3+7 q^4+10 q^5+16 q^6+\cdots\right)\\
q^{-17/26}\left(q+q^2+2 q^3+3 q^4+6 q^5+7 q^6+12 q^7+\cdots\right) \\
-q^{-9/26}\left(q^2+q^4+q^5+2 q^6+2 q^7+\cdots\right)\qquad\qquad\qquad\\
-q^{-3/26}\left(q+q^2+2 q^3+2 q^4+4 q^5+6 q^6+9 q^7+\cdots\right)\\
q^{-25/26}\left(q+q^2+2 q^3+4 q^4+6 q^5+9 q^6+14 q^7+\cdots\right)\\
-q^{-23/26}\left(q+2 q^2+3 q^3+5 q^4+8 q^5+12 q^6+18 q^7+\cdots\right)
\end{pmatrix}
\end{equation}
For the above rank-seven vvmf, using Eq. \eqref{chitransform} we obtain the following $T$ and $S$ matrices:
\begin{equation}
\begin{aligned}
&T=\text{Diag}\begin{pmatrix}
e^{i \pi} ,& 
 e^{-\frac{i \pi }{13}}, &  e^{-i \pi\left(\frac{17}{13}\right)}, & e^{-i \pi\left(\frac{9 }{13}\right)} , & e^{-i \pi\left( \frac{3  }{13}\right)}, &  e^{\frac{i \pi }{13}} , & e^{i \pi\left( \frac{3  }{13}\right)}\end{pmatrix}\\
 S&=\frac{1}{\sqrt{13}} \tiny\begin{bmatrix}
  1 & -2 & 2 & -2 & 2 & -2 & 2 \\
 -1 & 2 \cos \left(\frac{2 \pi }{13}\right) & -2 \sin \left(\frac{5 \pi }{26}\right) & 2 \sin
   \left(\frac{\pi }{26}\right) & 2 \sin \left(\frac{3 \pi }{26}\right) & -2 \cos \left(\frac{3 \pi
   }{13}\right) & 2 \cos \left(\frac{\pi }{13}\right) \\
 1 & -2 \sin \left(\frac{5 \pi }{26}\right) & -2 \sin \left(\frac{3 \pi }{26}\right) & 2 \cos
   \left(\frac{\pi }{13}\right) & -2 \cos \left(\frac{3 \pi }{13}\right) & -2 \sin \left(\frac{\pi
   }{26}\right) & 2 \cos \left(\frac{2 \pi }{13}\right) \\
 -1 & 2 \sin \left(\frac{\pi }{26}\right) & 2 \cos \left(\frac{\pi }{13}\right) & -2 \sin
   \left(\frac{3 \pi }{26}\right) & -2 \cos \left(\frac{2 \pi }{13}\right) & 2 \sin \left(\frac{5 \pi
   }{26}\right) & 2 \cos \left(\frac{3 \pi }{13}\right) \\
 1 & 2 \sin \left(\frac{3 \pi }{26}\right) & -2 \cos \left(\frac{3 \pi }{13}\right) & -2 \cos
   \left(\frac{2 \pi }{13}\right) & 2 \sin \left(\frac{\pi }{26}\right) & 2 \cos \left(\frac{\pi
   }{13}\right) & 2 \sin \left(\frac{5 \pi }{26}\right) \\
 -1 & -2 \cos \left(\frac{3 \pi }{13}\right) & -2 \sin \left(\frac{\pi }{26}\right) & 2 \sin
   \left(\frac{5 \pi }{26}\right) & 2 \cos \left(\frac{\pi }{13}\right) & 2 \cos \left(\frac{2 \pi
   }{13}\right) & 2 \sin \left(\frac{3 \pi }{26}\right) \\
 1 & 2 \cos \left(\frac{\pi }{13}\right) & 2 \cos \left(\frac{2 \pi }{13}\right) & 2 \cos
   \left(\frac{3 \pi }{13}\right) & 2 \sin \left(\frac{5 \pi }{26}\right) & 2 \sin \left(\frac{3 \pi
   }{26}\right) & 2 \sin \left(\frac{\pi }{26}\right) \notag
\end{bmatrix} 
\end{aligned}
\end{equation}

\section{Matrix differential equations and vvmfs}

The decomposition of Umbral Jacobi forms in terms of $\widehat{sl(2)}$ characters has given us vvmfs that are weight zero and have multiplier determined by the matrices $S$ and $T$. We have obtained the first few terms in their Fourier expansions by direct computation. In this section, we will determine them to all orders. For $N\neq 6$, we show that they are solutions to a Matrix Differential Equation (MDE) thereby obtaining explicit analytical formulae for the vvmfs. For $N=6$, we use a different method to obtain a similar result.

Let $\mathcal{M}^{!}_w(\rho)$ denote the space of weakly holomorphic vvmf  with multiplier $\rho$ of weight $w$ and rank $d$. Further, let 
\[
S=\rho\begin{pmatrix}
0 & 1 \\ -1 & 0
\end{pmatrix} \quad, \quad 
T=\rho\begin{pmatrix}
1 & 1 \\ 0 & 1
\end{pmatrix} \quad,\quad
U=\rho\begin{pmatrix}
0 & -1 \\ 1 & -1 
\end{pmatrix}=S T^{-1}\ .
\]
For $j=0,1$, let $a_j$ denote the multiplicity of the eigenvalue $(-1)^j$ of $S$ and for $j=0,1,2$ let $b_j$ denote the multiplicity of the eigenvalue $\exp(2\pi ij/3)$ of $U$. Further, let us assume that $T$ is diagonal 
\begin{equation}
T = \exp (2\pi i \Lambda) \quad,\quad \text{where } \Lambda =\text{Diag}(\lambda_1,\lambda_2,\ldots,\lambda_d)\ .
\end{equation}
The exponents $\lambda_i$ are only defined modulo one. In some situations, the exponents can be fixed. Let $\mathbf{g}(\tau)\in \mathcal{M}^{!}_w(\rho)$ have the following Laurent series
\begin{equation}
\mathbf{g}(\tau) = q^\Lambda\ \sum_{n\in \BZ}\mathbf{a}_n\ q^n\ .
\end{equation}
Define the principal part map, $\mathcal{P}_\Lambda: \mathcal{M}^{!}_w(\rho) \rightarrow \BC^d[q^{-1}]$ as follows:
\begin{equation}
\mathcal{P}_\Lambda(\mathbf{g}) =  \sum_{n\leq0} \mathbf{a}_n\ q^n\ .
\end{equation}
\begin{prop}[Gannon\cite{Gannon:2013jua}] Let $(\rho,w)$ be admissible and $T$ diagonal. Then,
there exists a choice of exponents $\Lambda$ for which the principal part map $\mathcal{P}_\Lambda : \mathcal{M}^{!}_w(\rho) \rightarrow \BC^d[q^{-1}]$  is a vector space isomorphism. 
\end{prop}
Using an index theorem argument, Gannon shows that a necessary but not sufficient condition for the bijectivity described above is 
\begin{equation}
\sum_{k=1}^d \lambda_k = c(\rho,w)\ ,
\end{equation}
where $c(\rho,w) := \frac{wd}{12}-\frac{a_1}{2}-\frac{b_1+2b_2}{3}$. In all our examples, we made choices that satisfied the above condition and for $N\leq 4$ found choices such that the bijection holds.

\begin{theorem}[Theorem 3.3(b) of  Gannon\cite{Gannon:2013jua}]
Let $(\rho,w)$ be admissible with rank $d$, $T$ diagonal and $\Lambda$ be bijective. Further,
let 
\[
\Xi(\tau):=\big(\mathbf{g}_1(\tau),\mathbf{g}_2(\tau),\ldots,\mathbf{g}_d(\tau)\big) = q^\Lambda\ (\mathbf{1}_d + \chi\ q + O(q^2))
\]
 denote the $d\times d$ matrix whose columns are a basis for $\mathcal{M}^{!}_w(\rho)$. Then, $\Xi(\tau)$, solves the Matrix Differential Equation (MDE) of the form:
\begin{equation}
\nabla_{1,w}\ \Xi(\tau) = \Xi(\tau)\Big((J(\tau)-984)\ \Lambda_w +\chi_w +[\Lambda_w,\chi_w]\Big)\quad,
\end{equation}
where $\Lambda_w = \Lambda - \frac{w}{12} \mathbf{1}_d$ and $\chi_w = \chi + 2w \mathbf{1}_d$. 
\end{theorem}
In all our examples, one column of $\Xi(\tau)$ is obtained from the vvmf that we obtain in Sec. \ref{decomposition} from the $\slhat$ decomposition of the Umbral Jacobi forms. We use the Fourier coefficents of the known vvmf  to determine the MDE.

\subsection{Identifying the MDE for vvmfs of interest}

The data entering the MDE of Gannon are the following:
\begin{enumerate}
\item The pair $(\rho,w)$,
\item an invertible set of exponents $\Lambda$, and
\item the $d\times d$ matrix $\chi$ defined by
\begin{equation}
\Xi(\tau) = q^\Lambda\ \Big( \mathbf{1}_d + \chi\ q + O(q^2)\Big)\ .
\end{equation}
\end{enumerate}
For all our examples, the weight $w=0$ and the multiplier $\rho$ is known. The unknowns are an invertible $\Lambda$ and  $\chi$.  Instead we know a solution to the MDE to any order that we desire. We assume that our solution corresponds to one column of $\Xi(\tau)$ -- this leads to a choice of $\Lambda$ and determines one column of $\chi$. We thus have $(d^2-d)$ unknowns that we determine by using higher orders of the known solution. This method works for $N=1,2,3$ but not for $N=6$ in part due to the higher dimensionality of the problem. There are no vvmfs for $N=4$ as the coefficients are constants and there is nothing more to do.

\subsubsection{$N=1$}

This is one of rank two and thus there are only two unknown constants to fix. We choose $\Lambda=\text{Diag}(-1/4,-11/12)$ and obtain
\begin{equation}
\chi= \begin{pmatrix}
 -84 & 32076 \\
 9 & 88 \\
\end{pmatrix} \ .
\end{equation}
This first column  agrees with the $O(q)$ term in $g_1(\tau)$ and  $g_2(\tau)$.

The expression for the full vvmf can be expressed in terms of the hypergeometric function\cite{Bantay:2007}
\begin{equation}
\Xi(\tau)=
\begin{pmatrix}
f(-1/4,7/6;z) & 32076 f(3/4,7/6;z)\\
9 f(1/12,7/6;z) & f(-11/12,7/6;z)
\end{pmatrix}\ ,
\end{equation}
where $z(\tau)=J(\tau)/1728$ and
\[
f(a,c;z) = (1728 z)^{-a} \ \ {}_2F_1(a, a+2/3; 2a+c;z^{-1})\ .
\]
The first column of $\Xi(\tau)$ is our solution. Thus, we can obtain the $q$-series for the vvmf associated with $A^{(1)}$ to arbitrary order.

\subsubsection{$N=2$}

We have a rank three vvmf with the exponents $\lambda_1=1/2\mod \BZ$, $\lambda_2=-1/10\mod \BZ $ and $\lambda_3=1/10\mod \BZ $, We find that the following exponents lead to an invertible $\Lambda$.
\[
\lambda_1=-1/2\ , \ \lambda_2 =-1/10\ , \ \lambda_3=-9/10\ .
\]
The eigenvalues of $S$ are $(1,1,-1)$ and thus $a_0=2$, $a_1=1$. The eigenvalues of $U$ are $1,\exp(2\pi i/3), \exp(4\pi i/3)$ and thus
$b_0=b_1=b_2=1$. From this, we see that $\sum_i \lambda_i =-3/2=c(\rho,0)$ where $c(\rho,0)=-a_1/2 -(b_1+2b_2)/3$.

The complete solution is given by the data 
\begin{equation}
\chi =\begin{pmatrix}
 222 & -10 & 4590 \\
 -1275 & 3 & 42483 \\
 25 & 3 & 27 
\end{pmatrix}\ ,
\end{equation}
and
\begin{equation}
{\footnotesize
\Xi(\tau)=\begin{pmatrix}
G\left(\lambda_1,\lambda_2,\lambda_3; z(\tau)\right) &
G\left(\lambda_1+1,\lambda_2-1,\lambda_3; z(\tau)\right) &
G\left(\lambda_1+1,\lambda_2,\lambda_3-1; z(\tau)\right)\\
G\left(\lambda_2+1,\lambda_1-1,\lambda_3; z(\tau)\right) &
G\left(\lambda_2,\lambda_1,\lambda_3; z(\tau)\right) &
G\left(\lambda_2+1,\lambda_1,\lambda_3-1; z(\tau)\right) \\
 G\left(\lambda_3+1,\lambda_1-1,\lambda_2; z(\tau)\right) &
 G\left(\lambda_3+1,\lambda_1,\lambda_2-1; z(\tau)\right) &
 G\left(\lambda_3,\lambda_1,\lambda_2; z(\tau)\right) 
\end{pmatrix}
}\ ,
\end{equation}
where 
\[
G(a,b,c;z):= (1728 z)^{-a}\  {}_3F_2 \Big(a,a+1/3,a+2/3;  a-b,a-c;z^{-1}\Big) \ .
\]
The second column of $\Xi(\tau)$ is our vvmf and is expressed in terms of generalized hypergeometric functions. This is true only for ranks $\leq 3$.

\subsubsection{$N=3$}

This is a rank 4 case and hence we do not anticipate that the solution can be expressed in terms of generalized hypergeometric functions.  We choose the following exponents:
\[
\lambda_1 =-3/4\quad,\quad
\lambda_2 =-25/28\quad,\quad
\lambda_1 =-1/28\quad,\quad
\lambda_4=-9/28\quad .
\]
The multiplicity of eigenvalues of $S$ and $U$  are 
\[
a_0= a_1=2\quad,\quad b_0=2\quad,\quad b_1=b_2=1\ .
\]
The choice of exponents satisfies the condition $\sum_i \lambda_i = c(\rho,0)=-2$. 
We obtain
\begin{equation}
\chi=\begin{pmatrix}
-150 & 550 & -2 & 36 \\
 49 & 25 & 1 & 15 \\
 -10829 & 37400 & 2 & -104 \\
 2499 & 10625 & -1 & -117
\end{pmatrix}
\end{equation}
The solution to the matrix DE is the following
\[
q^\Lambda \left(
\begin{smallmatrix}
1-150 q-39249 q^2-1624394 q^3 & 550 q+248490 q^2+15046550
   q^3 & -2 q-2 q^2-4 q^3 & 36 q+918
   q^2+9284 q^3 \\
 49 q+20874 q^2+1007244 q^3 & 1+25 q+27625 q^2+1978625
   q^3 & q+q^2+3 q^3& 15 q+576 q^2+6183
   q^3 \\
 -10829 q-614754 q^2-14799078 q^3 & 37400 q+3220140
   q^2+106417025 q^3 & 1+2 q+3 q^2+5 q^3 &
   -104 q-1107 q^2-8181 q^3 \\
 2499 q+217854 q^2+6319628 q^3 & 10625 q+1485800 q^2+60356369
   q^3 & -q-q^3 & 1-117 q-1647 q^2-13461
   q^3
\end{smallmatrix}\right) +O\left(q^4\right)
\]
where the third column is the vvmf of interest.
We have checked that column three of the above matrix agrees with expressions for $(g_1,\ldots,g_4)$ to $O(q^{16})$. Thus, even though we do not have simple expression in terms of hypergeometric functions as before, we have identified the MDE that the vvmf satisfies. We can easily solve the recursion relation to obtain the $q$-series to fairly high orders.

\subsubsection{$N=6$}

We choose the exponents as follows:
\begin{equation}
\left( \lambda_1,\ldots,\lambda_7 \right)=\left( -\frac{1}{2},-\frac{1}{26},-\frac{17}{26},-\frac{9}{26},-\frac{3}{26},-\frac{25}{26},-\frac{23}{26} \right)
\end{equation}
with $\sum\limits_{i} \lambda_i=-\frac{7}{2}$. The multiplicity of eigenvalues of $S$ and $U$  are 
\[
a_0=4\quad,\quad a_1=3\quad,\quad b_0=3\quad,\quad b_1=b_2=2\ .
\]
Hence $c_{\rho,0}=-7/2=\sum_i \lambda_i$. However, we have not been able to determine whether the choice of exponents is bijective. The problem is the large number of constants that need to be determined using the data from the known vvmf. Using the action of $\nabla_{i,w}$ for $i=1,2,3$, we can generate three linear combinations of the solutions. This leaves us with 21 unknown constants and this  space is too large for us to solve on a computer. Hence we chose an alternate method to get an all orders formula for the vvmf that we discuss next.

\subsection{Determining an explicit formula for the $N=6$ vvmf}

We observe that the theta expansion of the Umbral Jacobi form takes a very simple form after dividing out by a factor of $\eta(\tau)$. 
\begin{align}
\psi_{0,6}(\tau,z) = \frac{1}{\eta(\tau)}\Big(&
\theta_{24,2}(\tau,z)+ \theta_{24,26}(\tau,z)+\theta_{24,22}(\tau,z)+\theta_{24,46}(\tau,z) \nonumber \\ 
&-\theta_{24,10}(\tau,z)-\theta_{24,34}(\tau,z)-\theta_{24,14}(\tau,z)-\theta_{24,38}(\tau,z)\Big) \\
= \frac{1}{\eta(\tau)}\Big(& \mathcal{M}_{24,2}(\tau,z) + \mathcal{M}_{24,22}(\tau,z)- \mathcal{M}_{24,10}(\tau,z)-\mathcal{M}_{24,14}(\tau,z\Big)\ , \label{psi06m}
\end{align}
where $\mathcal{M}_{k,m}(\tau,z)=\theta_{k,m}(\tau,z)+\theta_{k,-m}(\tau,z)$.
Kac and Peterson\cite[see section 5.5]{KacPeterson1984} express the characters of $\slhat$ in terms of theta functions that appear above. The transformation matrix is given by Hecke modular forms. Explicitly, one has
\begin{equation}
\chi_{k,\lambda}(\tau,z)= \sum_{\substack{0\leq n < 2m \\ n\equiv\lambda\text{ mod }2}} \frac{\mathcal{C}^{(k)}_{\lambda,n}(\tau)}{\eta(\tau)^3} \ \theta_{k,n}(\tau,z)\ ,
\end{equation}
where $\mathcal{C}^{(k)}_{\lambda,n}$ is defined in terms of  Hecke indefinite modular forms as follows:
\begin{equation}
\mathcal{C}^{(k)}_{\lambda,n}(\tau)= \sum_{\substack{(x,y)\in\BR^2 \\ -|x|< y \leq |x|\\ (x,y)\text{ or }\left(\frac12-x,\frac12+y\right)\in \left(\frac{\lambda+1}{2(k+2)},\frac{n}{2k}\right)+\BZ^2}} \text{sign}(x)\ q^{(k+2)x^2-ky^2} \ .
\end{equation}
We need to express the theta functions in terms of $\slhat$ characters. This is given by
\begin{equation}\label{inverse}
\mathcal{M}_{k,n}(\tau,z)= \sum_{\substack{0\leq \lambda < 2k \\ \lambda\equiv n\text{ mod }2}} \mathcal{D}^{(k)}_{n,\lambda}(\tau) \ \chi_{k,\lambda}(\tau,z)\ ,
\end{equation}
where
\begin{equation}
\mathcal{D}^{(k)}_{n,\lambda}(\tau) = \sum_{\substack{m\in\BZ\\ m\equiv \pm n \text{ mod } 2k}} (-1)^{\frac{\lambda +m}2}\ q^{\frac{k(k+2)}{8}\left(\frac{m}{k}+\frac{\lambda+1}{k+2}\right)^2}\ .
\end{equation}
Substituting Eq. \eqref{inverse} in Eq. \eqref{psi06m}, we obtain
\begin{align}
\psi_{0,6}(\tau,z) &= \sum_{\lambda \equiv 0\text{ mod }2}\frac{\Big(\mathcal{D}^{(24)}_{2,\lambda}(\tau) + \mathcal{D}^{(24)}_{22,\lambda}(\tau)- \mathcal{D}^{(24)}_{10,\lambda}(\tau)-\mathcal{D}^{(24)}_{14,\lambda}(\tau)\Big)}{\eta(\tau)}\ \chi_{24,\lambda}(\tau,z) \notag \\
&= \sum_{\lambda \text{ even}}\left(\sum_{\substack{m\in\BZ\\ m \equiv \pm 10 \text{ mod } 24}}-\sum_{\substack{m\in\BZ\\ m\equiv\pm 2 \text{ mod } 24}}\right)\ \frac{(-1)^{\lambda/2}q^{78\left(\frac{m}{24}+\frac{\lambda+1}{26}\right)^2}}{\eta(\tau)}\ \chi_{24,\lambda}(\tau,z)
\end{align}
Comparing the above  expression with Eq. \eqref{psi06}, we obtain explicit formulae for $(g_1(\tau),\ldots,g_7(\tau))$ that agree with the expressions to the order that we have determined them.
We thus have obtained explicit formulae for the vvmf associated with $N=6$ even though we have not determined the MDE satisfied by the vvmf.

\subsection{Interpreting the vvmfs}

We have seen that the vvmf that we denote by $\mathbf{g}(\tau)$ captures the contribution of simple roots. Combining this result with the invariance of the Siegel modular forms under the action of the dihedral group, we obtain formulae that extend our results. The Fourier-Jacobi expansion of the Siegel modular form is compatible with the action of the subgroup $\langle w_2,\hdelta\,\rangle$ as these are realised as elements of the Jacobi group which preserve the cusp at $\tau'=i\infty$. The generator $\gamma^{(N)}$ does not belong to the Jacobi group. Including its action on the Umbral Jacobi form its decomposition into $\slhat$ characters enables us to organise the result in terms of orbits of the extended Weyl group.

The Siegel modular form $\Delta_{k(N)}(\mathbf{Z})$ can be written as a sum of terms of the kind that follow from our $\slhat$ decomposition of the Umbral Jacobi form.
\begin{enumerate}
\item The real roots $\alpha_0$ and $\alpha_3$ are accounted  from the expansion of
\[
\sum_{w\in W} \det(w)\ w\Big( e^{-\varrho^{(N)}}\Big)
\]
\item Let us denote the set of imaginary roots with zero norm $(\alpha_1+\alpha_2)$, $(\alpha_0 + \alpha_1)$, $(\alpha_2+\alpha_3)$ and their $\gamma^{(N)}$ translates  by $S_0$, These appear in the expansion of a Borcherds correction of the form
\begin{equation}
\sum_{w\in W}  \det(w)\ \sum_{a\in S_0} \sum_{n=1}^\infty \sigma_N(n)\  w\Big( e^{-\varrho^{(N)}+n a }\Big)
\end{equation}
where $\sigma_N(n)$ is defined as follows:
\begin{equation}
\prod_{m=1}^\infty (1-q^m)^{3(4-N)/N} = 1+\sum_{n=1}^\infty\sigma_N(n)\ q^N\ .
\end{equation}
It is easy to see that $\sigma_4(n)=0$ for all $n>1$ and $\sigma_6(n)=p(n)$ where $p(n)$ is the number of partitions of the positive integer $n$.
\item For $N\leq 4$, all other terms correspond to imaginary simple roots and provide Borcherds correction terms. To see how to do this, consider a term of the form
\[
g(\tau)\ \chi_{\Lambda} (\tau,z) = \sum_{m=0}^\infty b(m)\, q^m\ \chi_{\Lambda}(\tau,z)\ .
\]
The $m$-th term in the above sum is associated with the $\slhat$ weight vector $(\Lambda+m\delta)$ with multiplicity $b(m)$. A $W$-covariant expression that accounts for these roots is
\begin{equation}
\sum_{w\in W} \det(w)\ w\left(\sum_{m=1}^\infty b(m)\ e^{-\varrho-\Lambda + m\delta }\right)\ .
\end{equation}
\item The case of $N=6$ needs special attention. First, we get new simple real roots that we denoted by $\widetilde{\alpha}_1$ and $\widetilde{\alpha}_2$. The Siegel modular form is not invariant under Weyl reflections generated by these roots. Further $\langle \rho^{(N)},\widetilde{\alpha}_i\rangle =+1$ and not equal to $-1$. The multiplicity on the product side is $-1$ and hence they are fermionic roots. It appears that the term that we obtain arises as follows:
\begin{equation}
\sum_{w\in W} \det(w)\ w\left( \frac{e^{-\varrho^{(N)}}}{(1-e^{-\widetilde{\alpha}_i})}\right) =  \sum_{w\in W} \det(w)\ w\left( e^{-\varrho^{(N)}}(1+ e^{-\widetilde{\alpha}_i}+\cdots )\right)
\end{equation}
The above formula is conjectural as we have not checked if the pieces indicated by the ellipsis do appear. We also see that further imaginary roots involving the tilde roots also appear. They are of the form $(\widetilde{\alpha}_i+\alpha_j)$ for $i,j=1,2$. These do not appear in the set of positive roots that we obtain from the product formula. This is also true for the weights associated with labels 5 and 7. There is a cancellation of the form $1-1=0$. We can see a similar cancellation  in the WKB denominator formula for $\mathcal{B}_6(\slhat)$. The root $\delta=(\alpha_1+\alpha_2)$ does not appear on the product side. This is because this root appears as a non-simple bosonic imaginary root as well as a fermionic imaginary simple root (with the same weight). This suggests that the positive roots given by the product formula is incomplete and we need to take into account cancellations that occur. Our decomposition in terms of $\slhat$ characters is able to account for this.
\end{enumerate}

\section{Concluding Remarks}

The main result of this paper is a preliminary study of the WKB superdenominator formulae associated with BKM Lie superalgebras using a $\slhat$ subalgebra (and its Borcherds extension). In the current paper, we have restricted our study to include the first two additional simple real roots (and corresponding imaginary simple roots) that appear in the first Fourier-Jacobi coefficients of the Siegel modular forms. This leads to an interesting connection with vector-valued modular forms associated with some Umbral Jacobi forms. In all cases, we obtained relatively simple formulae for the Fourier coefficients of the vvmfs. These Fourier coefficients correspond to the multiplicities of simple roots, both imaginary and real, of the BKM Lie superalgebras.

The next step would be to carry out a similar decomposition for all Fourier-Jacobi coefficients. The connection with umbral moonshine gives a second formula for the Siegel modular forms. Extending heuristic arguments given in \cite{Govindarajan:2011em} (see also \cite{Dijkgraaf:1996xw}) for Mathieu moonshine to Umbral moonshine, one has
\begin{equation}
\Delta_{k(N)}(\mathbf{Z}):= s^{1/2}\ \phi_{k(N),1/2}(\tau,z) \ \exp\left[-\sum_{m=1}^\infty s^{mN} \ \psi_{0,N}~\Big|V_m~(\tau,z)\right]\ ,
\end{equation}
where 
\begin{equation}
 \psi_{0,N}~\Big|V_m~(\tau,z) = \frac1m \sum_{ad =m}\sum_{b=0}^{d-1}  \ \psi_{0,N}\left(\tfrac{a\tau+b}{d},az\right)\ . 
\end{equation}
The same formula also appears in \cite[see Eq. (2.7)]{GritsenkoNikulinII}. This formula is very useful in obtaining explicit formulae for higher Fourier-Jacobi coefficients of the $\Delta_{k(N)}(\mathbf{Z})$. For instance, the second coefficient is given by
\begin{equation}
\psi_{0,2N}(\tau,z) = \frac{1}{2} \left(\psi_{0,N}(\tau,z)\right)^2 - \sum_{ad=2} \sum_{b=0}^1 \psi_{0,N}\left(\tfrac{a\tau+b}{d},az\right)\ .
\end{equation}
The above formula has a nice interpretation. Let $\mathcal{V}_N$ denote a $\slhat$ module such that ($H$ is the Cartan subalgebra of $\slhat$)
\[
\mathcal{V}_N = \oplus_{\mu\in H} V_\mu\ , 
\] 
and the Umbral Jacobi form is equal to supercharacter of $\mathcal{V}_N$ i.e.,
\[
s^N\,\psi_{0,N}(\tau,z) = \text{Sch}(\mathcal{V}_N) := \sum_{\lambda\in H}  \Big(\text{dim}V_{0\,\mu} - \text{dim}V_{1\,\mu}\Big)\ e^{-\mu}\ .
\]
where $V_{0\,\mu}$ (resp. $V_{1\,\mu}$) is the bosonic (resp. fermionic) subspace of $\mathcal{V}_N$ of weight $\mu$.  Then, $\psi_{0,2N}(\tau,z)$ is obtained as the supertrace over direct sum of the $\slhat$ modules: $\Lambda^2\mathcal{V}_N$ and $\mathcal{V}_N^{[2]}$. The latter module $\mathcal{V}_N^{[2]}$ is obtained via the following {\em scaling procedure} \cite{Frenkel:1982}. The Lie subalgebra $\widehat{sl(2)}^{[2]} = sl(2)\otimes\mathbb{C}[t^2,t^{-2}]\oplus \mathbb{C}\,\hat{k} \oplus \mathbb{C}\, d $ of $\widehat{sl(2)}$ is in fact isomorphic to $\widehat{sl(2)}. $\footnote{via the isomorphism $X \otimes t^{2m} \mapsto X \otimes t^m$ for all $X \in {sl(2)}, m \in \mathbb{Z}$ and $\hat{k} \mapsto \hat{k}/2, \, d \mapsto 2d$ ({\em cf.} \S\ref{sec:sl2hatdef}).} The $\widehat{sl(2)}$-module $\mathcal{V}_N$ is $\mathbb{Z}_+$-graded, with the highest weight state being of grade zero and each application of $X \otimes t^{-m}$ increasing the grade by $m$. The subspace of $\mathcal{V}_N$ comprising its graded pieces of even grade is a module for the subalgebra $\widehat{sl(2)}^{[2]} \cong \widehat{sl(2)}$. This module is denoted $\mathcal{V}_N^{[2]}$. It is easy to see that 
\[
\text{Sch}\left(\mathcal{V}_N^{[2]}\right) =  s^{2N}\ \sum_{b=0}^1 \psi_{0,N}\left(\tfrac{\tau+b}{2},z\right)\ .
\]
Formulae such as these will enable us to write explicit formulae using the $\slhat$ decomposition obtained in this paper. This should, in principle, enable us to rewrite the sum side of the WKB denominator formula first in terms of $\slhat$ representations and then in sums where the covariance under the full Weyl group is manifest. We hope to report on this in the future\cite{WorkinProgress}.

In \cite{GritsenkoNikulinII}, Gritsenko and Nikulin point out that the $\Delta_{k(N)}(\mathbf{Z})$ for $N=1,2,3,4$ are three-dimensional generalizations of the Dedekind eta function. Rankin\cite{Rankin:1956} showed that the weight-twelve modular form 
$\Psi = \eta(\tau)^{24}$ of $\Gamma_1$ satisfies the following nonlinear ODE: (see Zagier\cite{Zagier:2008} for a derivation)
\begin{equation}
\label{RankinODE}
13 \Psi_1^4 + 10 \Psi^2 \Psi_1\Psi_3 - 24 \Psi \Psi_1^2 \Psi_2 +3 \Psi^2\Psi_2^2-2 \Psi^3\Psi_4=0\ ,
\end{equation}
where $\Psi_p\equiv \frac{d^p\Psi}{d\tau^p}$. Defining
\begin{equation}
y=\frac12 \frac{d}{d\tau} \log \Psi = \frac12 \frac{\Psi_1}{\Psi}\ ,
\end{equation}
Rankin's ODE becomes the Chazy equation:
\begin{equation}
\label{Chazy}
y''' -2y y''+3(y')^2 = 0\ .
\end{equation}
This nonlinear equation satisfies the Painlev\'e  property and connections with integrable systems (see \cite{Ablowitz:2003,Ashok:2018} and references therein).  
We have found MDE's for vvmf's associated with the Umbral Jacobi forms. Do all these combine to give a nice three-dimensional modular ODE for the logarithm of the Siegel modular forms? In this context, it is known that the logarithmic derivatives of the genus two theta constants satisfy a system of equations.\cite{Ohyama:1996,Zudilin:2000}. These methods might help one obtain similar nonlinear modular differential equations for the Siegel modular forms.

\appendix
\section{Modular background}

In this appendix, we discuss the different kinds of automorphic forms that appear in the paper. In particular, for vector-valued modular forms, we follow the discussion of Gannon\cite{Gannon:2013jua}.

\subsection{Basic Group Theory}

Let $\mathbb{H}_1$ denote the upper half-plane and $\mathbb{H}_1^*=\mathbb{H}_1 \cup \mathbb{Q}\cup \{\infty\}$ denote the extended upper half-plane. The group $\Gamma^{(1)}:=SL(2,\BZ)$ acts on $\mathbb{H}_1$ as follows:
\begin{equation}
\gamma \cdot \tau := \frac{a\tau + b }{c\tau +d }\ ,\quad \gamma= \left(\begin{smallmatrix}
a & b \\ c & d 
\end{smallmatrix}\right) \in \Gamma^{(1)}\text{ and } \tau \in \mathbb{H}_1\ .
\end{equation}

Let $\mathbb{H}_2$ denote the upper half-space with coordinates $\mathbf{Z}=\left(\begin{smallmatrix} \tau & z \\[2pt] z & \tau' \end{smallmatrix}\right)$.
The group $Sp(4,\mathbb{Q})$ is the set of $4\times 4$ matrices, $M$, written 
in terms of four $2\times 2$ matrices $A,\, B,\, C,\, D$ with entries in $\mathbb{Q}$
as
$$
M=\begin{pmatrix}
   A   & B   \\[3pt]
    C  &  D
\end{pmatrix}\ ,
$$
satisfying $ A B^T = B A^T $, $ CD^T=D C^T $ and $ AD^T-BC^T=I $. 
This group acts naturally 
on the Siegel upper half space, $\mathbb{H}_2$, as
\begin{equation}
\mathbf{Z}=\left(\begin{smallmatrix} \tau & z \\[2pt] z & \tau' \end{smallmatrix}\right)
\longmapsto M\cdot \mathbf{Z}\equiv (A \mathbf{Z} + B) 
(C\mathbf{Z} + D)^{-1} \ .
\end{equation}

The paramodular group at paramodular level $t$ that we denote by $\Gamma_t$ is defined as follows (we follow \cite{Gritsenko:2008} for all definitions) (for $t\in \mathbb{Z}_{>0}$):
\begin{equation}
\Gamma_t = \left\{
\left(\begin{smallmatrix}   
* & *t & * & * \\[2pt] * & * & * & *t^{-1} \\[2pt] * & *t & * & * \\[2pt] *t & *t & *t & *  
\end{smallmatrix}\right)\in Sp(4,\mathbb{Q}),\ \textrm{all } * \in \BZ \right\}\ .
\end{equation}
When $t=1$, then $\Gamma_1=Sp(4,\BZ)\equiv \Gamma^{(2)}$ is the usual  symplectic group.

Let $\Gamma^+_t=\Gamma_t\cup \Gamma_t V_t$  a normal double extension of $\Gamma_t$ in $Sp(4,\mathbb{R})$ with
\begin{equation}\label{Vtdef}
V_t = \tfrac1{\sqrt{t}} \left(\begin{smallmatrix} 0 & t & 0 & 0 \\ 1 & 0 & 0& 0 \\
0 & 0 & 0 & 1 \\ 0 & 0 & t & 0 \end{smallmatrix}\right)\ ,
\end{equation}
with $\det(CZ+D)=-1$.
This acts on $\mathbb{H}_2$ as 
\begin{equation}
(\tau,z,\tau' ) \longrightarrow (t \tau', z, \tau/t)\ .
\end{equation}
The group $\Gamma^+_t$ is generated by $V_t$ and its parabolic subgroup
\begin{equation}
\Gamma_t^\infty = \left\{
\left(\begin{smallmatrix}   
* & 0 & * & * \\[2pt] * & 1 & * & *t^{-1} \\[2pt] * & 0& * & * \\[2pt] 0 & 0 & 0 & 1  
\end{smallmatrix}\right)\in \Gamma_t,\ \textrm{all } * \in \BZ \right\}\ .
\end{equation}
The Jacobi group  is defined by 
\begin{equation}\label{JacobiGroup}
\Gamma^J=\big(\Gamma_t^\infty \cap Sp(4,\BZ)\big)/{\pm \mathbf{1}_4} \simeq \Gamma^{(1)}\ltimes H(\BZ)\ .
\end{equation} 
The embedding of 
$\left(\begin{smallmatrix} a & b \\ c & d\end{smallmatrix}\right)
\in SL(2,\BZ)$ in $\Gamma_t$ is given by
\begin{equation}
\label{sl2embed}
\widetilde{\begin{pmatrix} a & b \\ c & d \end{pmatrix}}
\equiv \begin{pmatrix}
   a   &  0 & b & 0   \\
     0 & 1 & 0 & 0 \\
     c &  0 & d & 0 \\
     0 & 0 & 0 & 1  
\end{pmatrix} \ .
\end{equation}
The above matrix acts on $\BH_2$ as
\begin{equation}
(\tau,z,\tau') \longrightarrow \left(\frac{a \tau + b}{c\tau+d},\  
\frac{z}{c\tau+d},\  \tau'-\frac{c z^2}{c \tau+d}\right)\ ,
\end{equation}
with $\det(C\mathbf{Z} + D)=(c\tau +d)$. The Heisenberg group, 
$H(\BZ)$, is generated by $Sp(4,\BZ)$ matrices of the form
\begin{equation}
\label{sl2embedapp}
[\lambda, \mu,\kappa]\equiv \begin{pmatrix}
   1   &  0 & 0 & \mu   \\
    \lambda & 1 & \mu & \kappa \\
     0 &  0 & 1 & -\lambda \\
     0 & 0 & 0 & 1  
\end{pmatrix}
\qquad \textrm{with } \lambda, \mu, \kappa \in \BZ
\end{equation}
The above matrix acts on $\BH_2$ as
\begin{equation}
(\tau,z,\sigma) \longrightarrow \left(\tau,\ z+ \lambda \tau  + \mu,\  
\tau' + \lambda^2 \tau + 2 \lambda z + \lambda \mu +\kappa \right)\ ,
\end{equation}
with $\det(C\mathbf{Z} + D)=1$. It is easy to see that $\Gamma^J$ 
preserves the one-dimensional cusp at $\textrm{Im}(\tau')= \infty$.
\subsection{Modular forms}

\begin{mydef}
A modular form of weight $w$ and character $\chi:\Gamma^{(1)}\rightarrow \BC^*$ is a map $f:\mathbb{H}_1^* \rightarrow \mathbb{C}$ such that
 \[
  f(\gamma\cdot \tau) = (c\tau + d)^w \ \chi(\gamma)\ f(\tau)\ .
\]
for all $\gamma=\left(\begin{smallmatrix}
a & b \\ c & d 
\end{smallmatrix}\right)\in \Gamma^{(1)}$.
\end{mydef}
A holomorphic modular form is holomorphic on the extended upper half-plane while a weakly holomorphic modular form is holomorphic on the upper half-plane and meromorphic on the extended upper half-plane. For $k\in \BZ_{>0}$, define the Eisenstein series as follows:
\begin{equation}
E_{2k}(\tau) = 1+ \frac{2}{\zeta(1-2k)}  \sum_{n=1}^\infty \sigma_{2k-1}(n)\ q^n\ ,
\end{equation}
where $q=\exp(2\pi i \tau)$ and $\sigma_s(n) = \sum_{d|n} d^s$ is the divisor function. For $k>1$, the Eisenstein series are holomorphic modular forms of weight $2k$. For $k=1$, it is not a modular form but $E_2^*(\tau) = E_2(\tau) -\frac{3}{\pi \text{Im}(\tau)}$ is not holomorphic but is modular of weight 2.
The Dedekind eta function is defined by 
\[
\eta(\tau):= q^{1/24} \prod_{m=1}^\infty (1-q^m)\ .
\]
It is a modular form of weight $\tfrac12$ of a subgroup of $\Gamma^{(1)}$. The $q^{1/24}$ implies that under $T$, it picks up a phase that is a 24-th root of unity. Taking the 24th-power of the Dedekind eta function gives us a modular form of weight 12 called the Discriminant function
\[
 \eta(\tau)^{24} = q -24 q^2 + 252 q^3 +\cdots \ .
\]
The modular $J$ function defined below is a weakly holomorphic modular form of weight zero. 
$$J(\tau):=\frac{E_4(\tau)^3}{\eta(\tau)^{24}}=q^{-1}+744+196884q+21493760q^2+\cdots$$ 
The $J$ function bijectively maps $\Gamma^{(1)}\backslash\mathbb{H}_1$ to the complex sphere.
At special points, $J(\exp(2\pi i/3)=0$ and $J(i)=1728$. Define $z(\tau)=J(\tau)/1728$. Thus, $z(\exp(2\pi i/3)=0$ and $z(i)=1$.

\subsection{Vector-Valued Modular Forms}
\begin{mydef}
An admissible multiplier systems $(\rho,w)$ consists of $w\in \BC$ called the weight and map $\rho: \Gamma^{(1)}\rightarrow GL(d,\BC)$ called the multiplier, for some positive integer $d$, called the rank, such that the following holds:
\begin{enumerate}
\item[(i)] the associated automorphy factor (with $\gamma =\begin{pmatrix}
a & b \\ c &d 
\end{pmatrix}\in \Gamma^{(1)}$)
\[
\widetilde{\rho}_w(\gamma,\tau):=\rho(\gamma) (c\tau +d)^w
\]
satisfies, for all $\gamma_1,\gamma_2 \in \Gamma^{(1)}$,
\begin{equation}
\widetilde{\rho}_w(\gamma_1\gamma_2,\tau) = \widetilde{\rho}_w(\gamma_1,\gamma_2\cdot\tau)\ \widetilde{\rho}_w(\gamma_2,\tau)\ ,
\end{equation}
\item[(ii)] $\rho(\mathbf{1}_2)=e^{-\pi i w}\,\rho(-\mathbf{1}_2)=\mathbf{1}_d$, where $\mathbf{1}_d$ is the $d\times d$ identity matrix. 
\end{enumerate}
\end{mydef}

\begin{mydef} Let $(\rho,w)$ be an admissible multiplier system of rank $d$. 
A vector-valued modular form (vvmf) $\mathbf{g}(\tau) = (g_1,g_2,\ldots,g_d)^T$ (of weight $w$, multiplier $\rho$ and rankd $d$) is a map $\mathbb{H}_1\rightarrow \BC^d$ provided
\begin{equation}
\mathbf{g}(\gamma\cdot \tau) = \widetilde{\rho}_w(\gamma,\tau)\ \mathbf{g}(\tau)\ ,
\end{equation}
for all $\gamma\in \Gamma^{(1)}$, $\tau\in\mathbb{H}_1$ and each component $g_i(\tau)$ is meromorphic in $\mathbb{H}_1^*$.
\end{mydef}
Let $\mathcal{M}^{!}_w(\rho)$ denote the space of weakly holomorphic vvmf i.e., those which are holomorphic in $\mathbb{H}_1$.

\subsection{Modular Differential Operators}

Let $f$ be a modular form of weight $w$ and $D_w$ denote the modular derivative i.e.,
\begin{equation}
D_w f(\tau) := \left(\frac{1}{2\pi i} \frac{d}{d\tau} -\frac{w}{12}\ E_2(\tau)\right) f(\tau)\ .
\end{equation}
This maps a modular form of weight $w$ to a modular form of weight $(w+2)$. Consider the differential operators that don't change weight.
\begin{equation}
\nabla_{1,w} = \frac{E_4(\tau)E_6(\tau)}{\eta(\tau)^{24}} \, D_w\ ,\quad
\nabla_{2,w} = \frac{E_4(\tau)^2}{\eta(\tau)^{24}} \, D_w^2\ ,\quad
\nabla_{3,w} = \frac{E_6(\tau)}{\eta(\tau)^{24}} \, D_w^3\ .\
\end{equation}

\subsection{Jacobi forms}

\begin{mydef} A Jacobi form of weight $k$ and index $m$ of $\Gamma_1$, is a holomorphic function
\[
\phi: \mathbb{H}_1 \times \BC \rightarrow \BC
\]
which transforms as follows under the Jacobi Group $\Gamma^J\simeq \Gamma_1\ltimes H(\BZ)$:
\begin{align*}
\phi(\frac{a\tau+b}{c\tau+d}, \frac{z}{c\tau +d }) &= (c\tau+d)^k \ e^{\frac{2\pi i mcz^2}{c\tau+d}}\ \phi(\tau,z)\quad,\quad \begin{pmatrix}
a & b \\ c & d 
\end{pmatrix}\in \Gamma_1 \\
\phi(\tau,z + \lambda \tau + \mu) &=e^{-2\pi i m(\lambda^2 \tau + 2 \lambda z)}\ \phi(\tau,z)\quad,\quad (\lambda,\mu)\in \BZ^2 \ .
\end{align*}
\end{mydef}
\noindent\textbf{Remark:} Jacobi forms of subgroups of $\Gamma_1$ are obtained by replacing $\Gamma_1$ by the appropriate subgroup.

The symmetries $\tau\rightarrow \tau +1$ and $z\rightarrow z+1$ imply the Fourier expansion for a Jacobi form of index $m$
\begin{equation}
\phi(\tau,z) = \sum_{n,\ell\in \BZ} c(n,\ell)\ q^n r^\ell\ ,
\end{equation}
where $q=\exp(2\pi i \tau)$ and $r=\exp(2\pi i z)$. The transformation under elliptic transformations implies that $c(n,\ell)$ depends on the combinations $(4nm-\ell^2)$ and $\ell$. A \textit{holomorphic} Jacobi form is one where $c(n,\ell)=0$ unless $4nm-\ell^2\geq0$. A weak Jacobi form is one for which $c(n,\ell)=0$ unless  $n\geq0$. A \textit{nearly holomorphic} Jacobi form if there exists $x\in \BZ_{\geq0}$ such that $\eta(\tau)^{24 x}\phi(\tau,z)$ is a  Jacobi form.

\subsubsection*{Examples}
\noindent The genus-one theta functions are defined by
\begin{equation}
\theta\left[\genfrac{}{}{0pt}{}{a}{b}\right] \left(\tau,z\right)
=\sum_{\ell \in \BZ} 
q^{\frac12 (\ell+\frac{a}2)^2}\ 
r^{(\ell+\frac{a}2)}\ e^{i\pi \ell b}\ ,
\end{equation}
where $a,b\in (0,1)\mod 2$. We define $\vartheta_1 
\left(\tau,z\right)\equiv\theta\left[\genfrac{}{}{0pt}{}{1}{1}\right](\tau,z)$,
$\vartheta_2 
\left(\tau,z\right)\equiv\theta\left[\genfrac{}{}{0pt}{}{1}{0}\right] 
\left(z_1,z\right)$, $\vartheta_3 
\left(\tau,z\right)\equiv\theta\left[\genfrac{}{}{0pt}{}{0}{0}\right] 
\left(\tau,z\right)$ and $\vartheta_4 
\left(\tau,z\right)\equiv\theta\left[\genfrac{}{}{0pt}{}{0}{1}\right] 
\left(\tau,z\right)$. \\

The function $\frac{\vartheta_1(\tau,z)^2}{\eta(\tau)^6}$ is a Jacobi form of weight $-2$ and index $1$.
Let $f_i =\vartheta_i(\tau,z)/\vartheta_i(\tau,0)$ for $i\in\{2,3,4\}$. The Umbral Jacobi forms at lambency $\ell$ are weak Jacobi forms of weight zero and index $(\ell-1)$.
\begin{equation}\label{UJFlist}
\begin{split}
&\psi_{0,1}(\tau,z) = 4( f_2^2 + f_3^2 + f_4^2)=\left(\tfrac{1}{r}+10 +r\right) + \cdots,\\
&\psi_{0,2}(\tau,z) = 2(f_2^2 f_3^2 + f_3^2 f_4^2 + f_4^2 f_2^2)=\left(\tfrac{1}{r}+4 +r\right)+\cdots ,\\
& \psi_{0,3}(\tau,z) = 4 f_2^2 f_3^2 f_4^2=\left(\tfrac{1}{r}+2 +r\right)+\cdots,\\
& \psi_{0,4}(\tau,z)=\frac14\left(  \psi_{0,1}(\tau,z)\psi_{0,3}(\tau,z)  - ( \psi_{0,2}(\tau,z)  )^2\right)=\left(\tfrac{1}{r}+1 +r\right)+\cdots ,\\
&\psi_{0,6}(\tau,z) =\psi_{0,2}(\tau,z) \psi_{0,4}(\tau,z) - ( \psi_{0,3}(\tau,z)  )^2=\left(\tfrac{1}{r} +r\right)+\cdots ,\\
\end{split}
\end{equation}

\subsection{Classical Theta functions}

\begin{align}
\theta_{k,\lambda}(\tau,z)&=\sum\limits_{m \in \mathbb{Z}}q^{k(m+\frac{\lambda}{2k})^2}r^{k(m+\frac{\lambda}{2k})}\ . \quad \lambda \in \mathbb{Z}/2k \mathbb{Z}
\end{align}
This is a vector-valued Jacobi form of weight half and index $k/4$. Dividing by $\eta(\tau)$ makes  the weight to zero.
\begin{align}
\alpha_{k,\lambda}(\tau,z)&:=\frac{\theta_{k,\lambda}(\tau,z)}{\eta(\tau)}
\end{align}
Under the $T$ and $S$ modular transformations, the $\alpha_{k,\lambda}$ transform as follows:
\begin{equation}
\begin{split}
 \alpha_{k,\lambda}(\tau+1,z)&= e^{2\pi i\left( \frac{\lambda^2}{4k} -\frac{1}{24}\right)}\ \alpha_{k,\lambda}(\tau,z)\\
\alpha_{k,\lambda}\left( - \frac{1}{\tau},\frac{z}{\tau}\right) &= e^{2\pi i  \frac{k z^2}{4\tau}}\; \sum\limits_{\mu=0}^{2k-1} \frac{e^{2 \pi i \left(- \frac{\lambda \mu }{2k} \right)}}{\sqrt{2k}} \alpha_{k,\mu}(\tau,z)
\end{split}
\end{equation}

Below we define the normalized $\slhat$ characters  which have nice modular properties.
\begin{align}\label{alphatransform}
\chi_{k,\lambda}(\tau,z)&=\frac{\theta_{k+2,\lambda+1}(\tau,z)-\theta_{k+2,-\lambda-1}(\tau,z)}{\theta_{2,1}(\tau,z)-\theta_{2,-1}(\tau,z)}; \text{\;for\;} k,\lambda \in \mathbb{Z}_{\geq 0}\ , \lambda\leq k\ .
\end{align}
Under the $T$ and $S$ modular transformations, one has
\begin{equation}\label{chitransform}
\begin{split}
 \chi_{k,\lambda}(\tau+1,z)&= e^{2\pi i\left[ \frac{(\lambda+1)^2}{4(k+2)}-\frac{1}{8} \right] } \chi_{k,\lambda}(\tau,z)\\
\chi_{k,\lambda}\left( - \frac{1}{\tau},\frac{z}{\tau}\right) &= e^{2\pi i k \frac{z^2}{4\tau}} \left(\frac{2}{k+2} \right)^\frac{1}{2} \sum\limits_{\mu=0}^k \sin\left[ \tfrac{\pi(\lambda+1)(\mu+1)}{k+2} \right] \chi_{k,\mu}(\tau,z)
\end{split}
\end{equation}

\subsection{Siegel Modular Forms}

\begin{mydef}
A Siegel modular form of weight $k$ and character $v$ 
with respect to $\Gamma_t$ is a holomorphic function $F: \BH_2 
\rightarrow \BC$ satisfying
\begin{equation}
F(M\cdot \mathbf{Z}) = v(M)\ \det(C\mathbf{Z}+D)^k \ F(\mathbf{Z})\ ,
\end{equation} 
for all $\mathbf{Z}\in \BH_2$ and $M\in \Gamma_t$.
\end{mydef}
The Fourier expansion of a Siegel modular form (with trivial character) with respect to the variable $\tau'$ (also called the Fourier-Jacobi expansion)
\begin{equation}
F(\mathbf{Z}) = \sum_{m=0}^\infty \phi_m(\tau,z)\ s^{tm} \ ,
\end{equation}
where $s=\exp(2\pi i \tau')$.
For each $m$, $\phi_m(\tau,z)$ is a Jacobi form of weight $k$ and index $mt$. This can be understood by observing that the cusp at $\tau'=i\infty$ is preserved by the subgroup $\Gamma^J$ and studying their transformation under this subgroup. We refer to the first non-vanishing term in the above Fourier expansion as the \textit{zeroth Fourier-Jacobi coefficient} of the Siegel modular form.

The character of Siegel modular forms are determined in part by their transformation under the Jacobi group $\Gamma^J$.  Consider the Jacobi form  of weight $-1$ and index $\frac12$:
\[
\frac{\vartheta_1(\tau,z)}{\eta(\tau)^3}\ .
\]
This has trivial character under modular transformations and the following character
\begin{equation}\label{vhdef}
v_H([\lambda,\mu,\kappa]) = (-1)^{\lambda+\mu+\lambda\mu +\kappa}\ .
\end{equation}
Multiplying the above Jacobi form by modular form $f(\tau)$ of $\Gamma_1$ with character $\chi$ leads to another Jacobi form of index half with character $(\chi\times v_H)$. This data can be obtained from the zeroth Fourier-Jacobi coefficient of the Siegel modular form. We need to determine the character under the involution $V_t$ ($q\leftrightarrow s^t$) and $[0,0,\kappa/t]$ (for $t>1$).

\section{Supercharacter formula  for BKM Lie superalgebras}
\subsection{The superdenominator identity}
Let $\mathfrak{g}$ be a BKM Lie superalgebra. The Weyl-Kac-Borcherds superdenominator identity of $\mathfrak{g}$ has the form $\mathcal{S} = \mathcal{P}$ (sum equals product). We describe this in greater detail here, closely following \cite{UrmieRay}.

Let $\mathfrak{g} = \mathfrak{g}_0 \oplus \mathfrak{g}_1$ be the decomposition of $\mathfrak{g}$ into bosonic (even) and fermionic (odd) subspaces. For $p=0, 1$, let $L^+_p$ denote the set of positive roots of bosonic ($p=0$) or fermionic ($p=1$) type, and  let $m_p(\alpha) = \dim (\mathfrak{g}_p)_\alpha$ denote the multiplicity of the root $\alpha$ in the subspace of appropriate parity.

The product side is given by:
\begin{equation}\label{eq:prodside}
  \mathcal{P} = \frac{\prod_{\alpha \in L^+_0} (1 - e^{-\alpha})^{m_0(\alpha)}}{\prod_{\alpha \in L^+_1} (1 - e^{-\alpha})^{m_1(\alpha)}}
\end{equation}
To describe the sum side, let $\alpha_i \, (i \in I)$ denote the simple roots of $\mathfrak{g}$. Let $I^{re} = \{i \in I: \langle \alpha_i, \alpha_i \rangle > 0\}$ and $I^{im} = I\backslash I^{re}$ be the subsets of real and imaginary simple roots. The Weyl group $W$ of $\mathfrak{g}$ (when $\mathfrak{g}$ is infinite-dimensional) is the group generated by the simple reflections $w_{\alpha_i}$ for $i \in I^{re}$. We can also decompose $I = I_0 \cup I_1$ into the disjoint union of simple roots of bosonic and fermionic types. Consider the set $\mathcal{T}$ of all elements $\mu$ in the root lattice of $\mathfrak{g}$ which can be expressed as a finite sum $\mu = \sum_{i \in I} k_i \alpha_i$ satisfying the following conditions:
\begin{enumerate}
\item $k_i$ is a non-negative integer for all $i$.
\item $k_i =0$ for  $i \in I^{re}$.
\item If $k_i$ and $k_j$ are nonzero for some $i \neq j$, then $\langle \alpha_i, \alpha_j\rangle =0$.
\item $k_i = 1$, unless $\alpha_i$ is an isotropic fermionic simple root, i.e., $i \in I_1$ with $\langle \alpha_i, \alpha_i \rangle = 0$.
\end{enumerate}

Given $\mu = \sum_{i \in I} k_i \alpha_i \in \mathcal{T}$, let $k_0(\mu) = \sum_{i \in I_0} k_i$ and $\epsilon_0(\mu) = (-1)^{k_0(\mu)}$. We define the {\em Borcherds correction}
\[ T = \sum_{\mu \in \mathcal{T}} \epsilon_0(\mu) e^{-\mu} \]
Finally, the sum side $\mathcal{S}$ is given by:
\begin{equation} \label{eq:sumside}
  \mathcal{S} = e^{\rho} \sum_{w \in W}\det(w)\ w\left( e^{-\rho}\, T\right)
  \end{equation}
The superdenominator identity is the equality of \eqref{eq:sumside} and \eqref{eq:prodside}. \\

\noindent \textbf{An example:} Consider a situation where has $m$ distinct bosonic simple roots of weight $(\delta,2\delta,3\delta,\ldots)$ with $\langle \delta,\delta \rangle=0$. The Borcherds correction factor due to these imaginary simple roots takes the form
\[
T = \prod_{k=1}^{\infty} (1-e^{-k\delta})^m\ .
\]
A negative value for $m$ corresponds to isotropic fermionic simple roots. We will encounter such Borcherds extensions of $\slhat$ i.e., $\slhat$ with the addition of the imaginary simple roots of the form discussed above.

\subsection{The supercharacter formula}\label{sec:supercharacter}
More generally, one has the Weyl-Kac-Borcherds formula for the supercharacter of an irreducible integrable highest weight module $L(\Lambda)$ of $\mathfrak{g}$. Here $\Lambda$ is a dominant integral weight of $\mathfrak{g}$, i.e., $(\Lambda, \alpha_i)$ is a non-negative integer (resp. real number) for $i \in I^{re}$ (resp. $i \in I^{im}$). We define a subset $\mathcal{T}_\Lambda$ of $\mathcal{T}$ by imposing the following extra condition in addition to (1)-(4) above:

\begin{enumerate}
  \item[5.] $k_i=0$ if $\langle \Lambda, \alpha_i\rangle < 0$.
  \end{enumerate}
Analogous to the above, define $ \widetilde{T}_\Lambda = \sum_{\mu \in \mathcal{T}_\Lambda} \epsilon'(\mu) e^{-\mu} $ and  
\begin{equation*} 
  \mathcal{S}_\Lambda = e^{\rho} \sum_{w \in W}\det(w)\ w\left( e^{-\rho-\Lambda}\, \widetilde{T}_\Lambda\right).
  \end{equation*}
The WKB supercharacter formula states that the supercharacter $\chi_\Lambda$  of $L(\Lambda)$ is given by the quotient 
\begin{equation}\label{WeylCharacterFormula}
\text{Sch}(L(\lambda)):=\chi_\Lambda= \frac{\mathcal{S}_\Lambda }{ \mathcal{P}}\quad.
\end{equation}
 Since $L(\Lambda)$ is the one-dimensional trivial representation when $\Lambda=0$, this reduces to the superdenominator identity in that case\cite{UrmieRay}.
\bibliographystyle{utphys}
\bibliography{refs}
\end{document}